\documentclass[11pt]{article}
\usepackage{latexsym}
\usepackage{amssymb}
\usepackage{amsmath}
\usepackage{graphicx}
\usepackage[hidelinks]{hyperref}

\newtheorem{Theorem}{Theorem}[section]

\newtheorem{Proposition}[Theorem]{Proposition}
\newtheorem{Assumption}[Theorem]{Assumption}

\newtheorem{Corollary}[Theorem]{Corollary}
\newtheorem{Remark}[Theorem]{Remark}

\usepackage[usenames]{color}
\definecolor{red}{rgb}{1.0,0.0,0.0}

\definecolor{blu}{rgb}{0.0,0.0,1.0}

\def \R{\mathbb{R}}

\def \E{\mathbb{E}}

\def \P{\mathbb{P}}
\def \Q{\mathbb{Q}}

\def \Ac{{\cal A}}

\def \Fc{{\cal F}}

\def \ni{\noindent}

\def \eps{\varepsilon}

\long\def\symbolfootnote[#1]#2{\begingroup%
\def\thefootnote{\fnsymbol{footnote}}\footnote[#1]{#2}\endgroup}

\def\reff#1{{\rm(\ref{#1})}}

\def\beqs{\begin{eqnarray*}}
\def\enqs{\end{eqnarray*}}
\def\beq{\begin{eqnarray}}
\def\enq{\end{eqnarray}}

\date{}

\begin{document}


\title{Utility maximization  with current  utility on the  wealth: regularity of solutions to the HJB equation\,\footnote{This research has been partially supported by the PRIN research project ``Metodi deterministici e stocastici nello studio di problemi di evoluzione"  of the Italian Minister of University and Research. {The author thank two anonymous Referees and the Editors for careful scrutiny and useful suggestions that allowed to improve substantially the paper}.}}
\author{Salvatore Federico$^{a)}$  \and Paul Gassiat$^{b)}$ \and Fausto Gozzi$^{c)}$}
    \maketitle

\symbolfootnote[0]{$^{a)}$ Dipartimento di Economia, Management e Metodi Quantitativi, Universit\`a di Milano, Italy. E-mail: \texttt{salvatore.federico@unimi.it}}
\symbolfootnote[0]{$^{b)}$ Institut f\"ur Mathematik, TU Berlin. E-mail: \texttt{gassiat@math.tu-berlin.de}}
\symbolfootnote[0]{$^{c)}$ Dipartimento di Economia e Finanza, Libera Universit\`a degli Studi Sociali ``Guido Carli'', Roma, Italy. E-mail:
    \texttt{fgozzi@luiss.it}}

\begin{abstract}
{ This paper deals with an  investment-consumption portfolio problem  when the current utility depends also on the wealth process. Such kind of problems arise, e.g., in portfolio optimization with random horizon or with random trading times. To overcome the difficulties of the problem a dual approach is employed: a dual control problem is defined and  treated by means of dynamic programming, showing that the viscosity solutions of the associated Hamilton-Jacobi-Bellman equation belong to a suitable class of  smooth functions. This allows to define a smooth solution of the primal Hamilton-Jacobi-Bellman equation and to prove, by verification,  that such solution is indeed unique in a suitable class of smooth functions and coincides with the value function of the primal problem.
Applications of the results to specific financial problems are given.
}
\end{abstract}


\bigskip


\tableofcontents

\section{Introduction}
This paper deals with the problem of utility maximization in consumption-investment models over a fixed horizon  when the current utility depends also on the wealth process. The fact that the current utility may depend also on the wealth is motivated by the fact that this situation arises in some concrete financial problems, as discussed in Section \ref{sec:app}.\medskip\\
We tackle the problem by duality  and using a dynamic programming approach  both on the primal and on the dual problem. Since the papers  by Karatzas, Lehoczky and Shreve
\cite{KLS87} and by Cox and Huang \cite{CH89}, the duality approach to consumption-investment problems has been extensively treated in the literature  (see the survey paper by Rogers \cite{R}, and  the book by Karatzas and Shreve \cite[Ch.\,3\ and 6]{KS} - and the references therein) to treat generalizations of the classical Merton problem  (incomplete markets, non-Markovian setting,  strategies constraints, transaction costs,  etc.). Notably with regard to our paper,  Bouchard and Pham \cite{BP04} treat  the case of current utility depending on the wealth in a semimartingale setting without developing the dynamic programming approach.

When the stock is assumed to evolve according to a stochastic differential equation, one can apply the dynamic programming machinery both to the primal and the dual problem to get some more insights on the solution of the problem. In particular the duality can be read at the analytical level of the Hamilton-Jacobi-Bellman (HJB) equation, providing a dual equation. This is what is done in Bian, Miao and Zheng \cite{BMZ} (see also the extension of such results in \cite{BZ}) in the case of no current utility on the wealth. But, as far as we know, duality has been never employed  combined with the dynamic programming when the current utility depends on the wealth process.
This  may be due to  the fact that when there is no dependence of the current utility on the wealth process the HJB equation associated to the  dual problem is linear - so approachable by semi-explicit solution written in terms of the heat kernel (see \cite{BMZ, BZ}) - while when the current utility also depends on the wealth such HJB equation is just semi-linear - so more difficult to study. At the level of control problems, this corresponds to the fact that in the former case the dual problem is simpler, as the control does not appear in it,    while in the latter one the dual problem is a real control problem   (these issues are discussed in Remark \ref{rem:dual}). Nevertheless, also  in this last case, the dual control problem is still simpler to treat than the primal one, as the control only appears in the drift of the process, consistently with the fact that the HJB equation is semilinear (while the HJB equation associated to the primal control problem is fully nonlinear and degenerate, so very difficult to tackle directly by the PDE's theory of classical solutions).\footnote{We also should mention the paper \cite{ST04}, where the HJB equation associated to the dual problem is again fully nonlinear, but admits a semi-explicit solution in the form of a power series.}   \medskip\\
Our method to solve the problem is the  following.


\emph{Step 1}: Starting from the original primal problem (with value function $V$ and an associated  primal HJB equation),  we  define a  dual problem, which is still a control problem,

\emph{Step 2}: We associate to the dual problem a  dual HJB equation and prove that  the value function $W$ of this dual problem is a viscosity solution of the dual HJB equation (Proposition \ref{propDualVisc}).

\emph{Step 3}:
 Since the dual HJB equation is semilinear and nondegenerate, we are able to prove good regularity results for $W$. This is proved in Theorem \ref{prop:regpsic12}, which is the key result of the paper.

\emph{Step 4}: The regularity of $W$ allows to define a smooth solution to the primal HJB equation, which is  the Legendre transform $\widetilde{W}$ of $W$.

\emph{Step 5}: We prove a verification theorem for our primal problem within a suitable class $\mathcal{C}$ of smooth  solutions of the primal HJB equation. Since $\widetilde{W}\in\mathcal{C}$, this theorem, together with a result of existence and uniqueness for the associated closed loop equation, will imply that $\widetilde{W}=V$  and that $V$ is the unique classical solution of the primal  HJB equation within the class $\mathcal{C}$. These results will yield also the construction of an optimal feedback control for the primal problem.
 \medskip\\
The rest of the paper is organized as follows. In Section 2 we set the problem and state the assumptions. In Section 3 we define the dual problem
 (Step 1 above). In Section 4 we  study the dual HJB equation by a viscosity approach  and  state the regularity of the value function $W$ (Steps 2 and 3 above). In Section 5 we prove that $V$ is a classical solution of the HJB equation and provide the optimal feedbacks through a verification theorem (Steps 4 and  5 above); {moreover we also provide an alternative approach based on the exploiting of the duality at a probabilistic level.} Finally,  Section 6 provides two concrete applications of our framework.

\section{Model and optimal control problem}

In this section we present the financial model and the (primal) stochastic control problem we deal with.
\smallskip\\
Let us consider a complete filtered probability space $(\Omega,\Fc,(\Fc_t)_{t\geq 0},\P)$ satisfying the usual conditions, on which is defined
a standard Brownian motion $(B_t)_{t\geq 0}$. We assume that $(\Fc_t)_{t\geq 0}$ is the filtration generated by this Brownian motion and enlarged by the $\P$-null sets.

On this space we consider a riskless asset with deterministic rate of return that without loss of generality (see Remark \ref{rem:negp}(ii) below) we set equal to $0$,  and a risky asset $S=(S_t)_{t\geq 0}$ with dynamics
%
$$dS_t \ =\  S_t\,( b(t) dt + \sigma(t) dB_t),$$
where $b, \sigma$ are deterministic coefficients representing, respectively, the drift and the volatility of the risky asset. \\\\
Fix a time horizon $T>0$.
In the setting above,
we define a set of admissible trading/consumption strategies in the following way.
Consider all the couples of processes $(c,\pi)$ such that
\begin{itemize}
	\item[(h1)] $c=(c_t)_{t\in[0,T]}$ is a real nonnegative process $(\Fc_t)_{t\in[0,T]}$-predictable and with  trajectories locally integrable in $[0,T)$; $c_t$ represents the consumption rate at time $t$;
	\item[(h2)]$\pi = (\pi_t)_{t\in[0,T]}$ is a  real  process $(\Fc_t)_{t\in[0,T]}$-predictable and with  trajectories  locally square integrable in $[0,T)$; $\pi_t$ represents the amount of money invested in the risky asset at time $t$.
	\end{itemize}
Given a couple  $(c,\pi)$ satisfying the requirements (h1)-(h2) above, we can consider the process $X_t$ representing the wealth associated to such strategy. Its dynamics are given by
\beq\label{wealth}
\begin{cases}
dX_t\ =\ \pi_t(b(t) dt+\sigma(t)dB_t)-c_t dt,\\
X_0\ =\ x_0,
\end{cases}
\enq
where $x_0\geq 0$ is the initial wealth.
As class of admissible controls we consider the couples  of processes $(c,\pi)$ satisfying (h1)-(h2) and such that the corresponding wealth process $X$ is nonnegative (no-bankruptcy constraint).
 The optimization problem is
\beq\label{eq:optprob}
\E \left[ \int_0^{T} U_1(t,c_t,X_t) dt+   U_2(X_T)\right].
\enq\\
{{We introduce the following notations that will be used in the paper.
\begin{itemize}
\item[-] $\R_+:=[0,+\infty)$.
\item[-] Given an integer $k\geq 0$, a real number $\delta\in(0,1]$ and $\mathcal{O}\subset \mathbb{R}^n$ open,
the symbol  $C_{loc}^{\frac{\delta}{2}, k + \delta}([0,T)\times \mathcal{O};\R)$ shall denote
   the space of real continuous functions on $[0,T) \times \mathcal{O}$ such that all the {space derivatives} up to order $k$ exist and are $\delta/2$-H\"older continuous with respect to $t$ and  $\delta$-H\"older continuous with respect to the space variables on each  compact subset of $[0,T) \times \mathcal{O}$.
   \item[-] Given an integer $k\geq 0$, a real number $\delta\in(0,1]$ and $\mathcal{O}\subset \mathbb{R}^n$ open,
the symbol  $C_{loc}^{1+\frac{\delta}{2}, k + \delta}([0,T)\times \mathcal{O};\R)$ shall denote
   the space of real continuous functions on $[0,T) \times \mathcal{O}$ such that the  first time derivative and all the {space derivatives} up to order $k$ exist and are $\delta/2$-H\"older continuous with respect to $t$ and  $\delta$-H\"older continuous with respect to the space variables on each  compact subset of $[0,T) \times \mathcal{O}$.
   \end{itemize}}
   \smallskip
   We make the following assumptions on the model.
\begin{Assumption}\label{ass:b,sigma}
$b, \sigma:[0,T]\to\R$ are strictly positive and $(\delta/2)$-H\"older continuous for some $\delta \in (0,1]$.
\end{Assumption}
\begin{Assumption}\label{ass:U}
The preference of the agent are described by  utility functions $U_1,U_2$ satisfying the following:
\begin{itemize}
\item[(i)]
$U_1: [0,T)\times \R_+^2\rightarrow \R$ is such that  $U_1\in C_{loc}^{\delta/2, k + \delta}([0,T) \times (0,+\infty)\times(0,+\infty);\R)$ for some $k\geq 2$ (and the same $\delta$ of Assumption \ref{ass:b,sigma}). For each fixed $t$ $\in$ $[0,T)$ the function $U_1(t,\cdot,\cdot)$ is  concave with respect to $(c,x)$ and  {nondecreasing} with respect to both the variables $c,x$.

 Moreover either
 $$(a) \ \ \ \begin{cases}
  \frac{\partial}{\partial c} \,U_1(t,0^+,x) = + \infty,\ \ \ \ \forall (t,x)\in[0,T)\times \R_+,\\\\
   \frac{\partial}{\partial c} \,U_1(t,+\infty,x) = 0,\ \ \ \ \forall (t,x)\in[0,T)\times \R_+,\\\\
   {\frac{\partial}{\partial c}\, U_1>0}, \  \ \frac{\partial^2}{\partial c^2}\, U_1< 0,
  \ \ \
  \   \mbox{in}\   [0,T) \times (0,+\infty)\times (0,+\infty),
  \end{cases}
 $$
 or
 $$(b) \ \ \ \frac{\partial}{\partial c}U_1\equiv 0.$$


\item[(ii)]
$U_2: \R_+\rightarrow \R$ is continuous,  {nondecreasing}, concave. Without loss of generality we assume
\beq\label{boundU_2}
U_2(0)&=&0.
\enq
\item[(iii)]
The following growth condition holds: there exist $K>0$ and $p\in(0,1)$ such that
\beq\label{growthU}
U_1(t,c,x) + U_2(x) \ \ \leq \ \ K (1+c^{p}+x^{p}), \ \ \  \forall (t,c,x)\in [0,T)\times \R_+^2.
\enq
Moreover, without loss of generality for the optimization problem, we assume that
 \beq\label{boundU_1}
 U_1(t,0,0)&=&0, \ \ \ \forall t\in[0,T).
 \enq

\item[(iv)] Either
$$ {(a) \ \ \exists\, \varepsilon >0 \ \mbox{such that} \ \lim_{c\rightarrow +\infty} U_1(t,c,0)\ =\ +\infty \ \ \mbox{uniformly in } t\in [T-\varepsilon,T),}$$
or
$$ (b) \  \ \lim_{x\rightarrow+\infty}U_2(x)\ \ =\ \ +\infty$$
or both.
\end{itemize}
\end{Assumption}
In the remark below we comment on some features of the model and explain when and how they can be eventually modified to cover other interesting cases.
\begin{Remark}\label{rem:negp}
\textbf{(i)}
In the applications one is often interested to work with power utility functions.  Assumption \ref{ass:U}
 includes only the case of positive power. On one hand the case of negative exponent
is interesting, as it seems to be even more realistic from the point of view of the agents' behavior; on the other hand,  it would require a slightly different treatment. Just for simplicity, we will work with Assumption \ref{ass:U},  nevertheless we stress that the case of negative power utility can be treated by the same techniques by suitable modifications.

\textbf{(ii)} The assumption that the riskless rate of return is $0$ can be done without loss of generality. Indeed, since  we are considering a quite general time-dependent $U_1$,  the interest rate can be discarded in it by a suitable discounting of the variables (see  \cite[Rem.\,2, p.\,189]{Kab}).

\textbf{(iii)}
The problem without consumption falls in our setting as well. Indeed,  take a problem without consumption and with running utility $u_1(t,x)$. Defining $U_1(t,c,x)=u_1(t,x)$ in our setting, consuming turns out to be not convenient, as  its negative effect on the wealth does not have a trade-off in terms of utility from consumption. In other terms, the optimal consumption is $c^*_t\equiv 0$. As a consequence the problem in our setting with $U_1$ defined as above is equivalent to the problem without consumption and with utility function $u_1$. In particular, when $u_1\equiv 0$ we fall in the setting of \cite{BMZ}.

\textbf{(iv)} We have set the problem with finite horizon. However, some problems arising in the applications - see Section \ref{sec:app} - involve the infinite horizon case, where $T={+\infty}$, for which the functional usually looks like
\beqs
\E \left[ \int_0^{{\infty}} e^{-\rho t} U_1(t,c_t,X_t) dt\right],
\enqs
where, as usual for infinite horizon problems,  $\rho>0$ is a discount rate sufficiently large to guarantee the finiteness of the value function.
The results we provide in the present paper for the finite horizon case can be suitably generalized to the infinite horizon case, with the complication of dealing in the viscosity treatment of the HJB equation with growth conditions for $t\rightarrow {+\infty}$ in place of terminal boundary conditions at $t=T$. We refer, e.g.,  to \cite{FG} for an example of the technical treatment of this kind of conditions and stress here that our main results - the regularity results - do not ``see" whether the horizon is finite or infinite, as they are based on local arguments. Of course, in this case one needs to assume that Assumption \ref{ass:U}(iv) is satisfied at point (a).

{\textbf{(v)} We comment on Assumption \ref{ass:U}(i). It  requires that either $U_1$ is independent of $c$ or it satisfies Inada's conditions with respect to $c$.
We need this assumption to get in a straightforward way the regularity of the Legendre transform of $U_1$ with respect to $c$ (Proposition \ref{prop:tildef}(6)), which is in turn needed to get the regularity of the dual value function, see Section \ref{sec:reg}.  Basically it is thought to cover the case of separable utility in the form $U_1(t,c,x)=U_1^{(1)}(t,c)\,+\,U_1^{(2)}(t,x)$, where $U_1^{(1)}$ is identically $0$ or satisfies the Inada conditions with respect to $c$, which is the case arising in the applications we have in mind (see Section \ref{sec:app}).
Relaxing  this assumption seems possible, but at a price of more demanding technical arguments.  We prefer to avoid such technicalities in order to focus on the main topic of the paper, which is the the regularity of solutions of the HJB equation by means of the duality approach.}

{\textbf{(vi)} The assumption of strict positivity of $b,\sigma$ is done to have strict parabolicity of the HJB equation. Actually this is needed only in the interior, so we might allow the cases $b(T)=0$ and/or $\sigma(T)=0$. However, allowing that would bring some other technicalities,
so we prefer to  impose strict positivity also at $T$.
We also stress that we  actually need just the assumption $b(t)\neq 0$ for all $t\in[0,T]$; but, due to continuity, this is equivalent to say  that $b$ keeps the sign. Since the assumption making sense from a financial point of view is $b(\cdot)>0$, we impose it.}

{\textbf{(vii)}
Although for simplicity we consider in our model the case of just one risky asset, it is easy to see that the program we described in the introduction  works also in more dimensions (more risky assets, as in \cite{BMZ}). In that case strict positivity of $b(t)$ and $\sigma(t)$ in Assumption \ref{ass:b,sigma} should be replaced by the assumption that for all $t\in[0,T)$ (the matrix) $\sigma(t)$ is invertible and (the vector) $b(t) \neq 0$, so that in the dual HJB equation \reff{eqDualHJB} the term $| \sigma^{-1}(t)b(t)|^2$ is then still well-defined and strictly positive.}

{\textbf{(viii)} We are concerned with a utility maximization problem. Nevertheless,  our approach seems applicable also to different cases, e.g. to the case of quadratic risk minimization, by suitably adapting the arguments.}

\end{Remark}

\section{Primal and dual control problem}\label{sec:primdual}
Since we are going  to apply the dynamic programming techniques, we define the optimization problem for generic initial data $(t,x)\in[0,T]\times \R_+$.
Let $t\in[0,T]$ and consider all the couples of processes $(c,\pi)$ such that
\begin{itemize}
	\item[(h1$'$)] $c=(c_s)_{s\in[t,T]}$ is a real nonnegative process $(\Fc_s)_{s\in[t,T]}$-predictable and with  trajectories locally integrable in $[0,T)$.
		\item[(h2$'$)]$\pi = (\pi_s)_{s\in[t,T]}$ is a  real  process $(\Fc_s)_{s\in[t,T]}$-predictable and with   trajectories square locally integrable in $[0,T)$.
			\end{itemize}
Given $x\geq 0$ and a couple  $(c,\pi)$ satisfying the requirements (h1$'$)-(h2$'$) above, we denote by $X^{t,x,c,\pi}$ the solution to \eqref{wealth} starting at time $t$ from $x$ and under the control $(c,\pi)$.
We define a class of admissible controls $\mathcal{A}(t,x)$ depending on the initial $(t,x)\in[0,T]\times[0,+\infty)$ as the set of couples $(c,\pi)$ satisfying the requirement above and such that the corresponding state trajectory $X^{t,x,c,\pi}$ is nonnegative. We notice that such set is nonempty for each $t\in[0,T]$ and $x\geq 0$, as for such initial data the null strategy $(c,\pi)\equiv(0,0)$ is always admissible. Moreover $\mathcal{A}(t,x)=\{(0,0)\}$ if and only if $x=0$.
Then  we define the functional
$$J(t,x;c,\pi)\ :=\ \E \left[ \int_t^{T}  U_1(s,c_s,X^{t,x,c,\pi}_s) ds+  U_2(X_T^{t,x,c,\pi})\right].$$
We call \emph{primal control problem} - and denote it by \textbf{(P)} - the optimization problem
\beqs
\mbox{\textbf{(P)}} \ \ \ \ \ \  \sup_{(c,\pi) \in \mathcal{A}(t,x)}J(t,x;c,\pi),
\enqs
and denote by $V$ the value function associated to this problem - that we call \emph{primal value function}, i.e.
\beqs
V(t,x)&:=&\sup_{(c,\pi) \in \mathcal{A}(t,x)}J(t,x;c,\pi), \ \ \ \ (t,x)\in[0,T]\times\R_+.
\enqs
Due to the fact that the state $0$ is an absorbing boundary for the problem and to \eqref{boundU_1}-\eqref{boundU_2}, we see that  $V$ satisfies the boundary condition
\beq \label{Bndryv}
{V}(t,0)\ = \ 0, \ \ \ \ \forall t\in[0,T].
\enq
On the other hand $V$ clearly satisfies also the the terminal condition
\beq\label{term}
V(T,x)&=&U_2(x).
\enq
Set
$$D_T\ :=\ [0,T)\times (0,{+\infty}).$$
By standard arguments of stochastic control (see e.g. \cite[Ch.\,4]{YZ}), we can associate to ${V}$ a HJB equation in $D_T$, which we call \emph{primal} HJB equation. It is
\beq\label{HJBv}
- {v}_t(t,x)   - \sup_{c\geq 0,\, \pi\in\R} H_{cv}(t,x,v_x(t,x),v_{xx}(t,x);c,\pi)
 &=& 0,
\enq
where the function $H_{cv}$ is defined for $(t,x,y,Q) \in D_T \times \R^2$, $ c\geq 0, \pi\in \R$, as
\beqs
H_{cv}(t,x,y,Q;c,\pi) &:=&  U_1(t,c,x) +(b(t) \pi- c) y +  \frac{\sigma(t)^2}{2} \pi^2 Q.
\enqs
When $y >0$ and $Q<0$  (the case we shall consider), the Hamiltonian
$$H(t,x,y,Q) \ \ :=\ \ \sup_{c\geq 0, \, \pi\in\R} H_{cv}(t,x,y,Q;c,\pi)$$
is finite and takes the form
\beq\label{hamiltspecial}
H(t,x,y,Q)  &=& {U}^*_1(t,y,x)- \frac{b^2(t)}{2\sigma^2(t)}\frac{y^2}{Q},
\enq
where $U_1^*$ is the sup-Legendre transform of $U_1$ with respect to $c$, i.e. the function (convex in $y$)
$$
 {U}^*_1(t,y,x)\ \ :=\ \ \sup_{c\geq 0}\  \{U_1(t,c,x)-cy\}, \ \ \  \ (t,y,x)\in [0,T)\times (0,+\infty)\times \R_+.
$$
We expect that $V$ may be characterized as solution of \eqref{HJBv} completed by the boundary and terminal conditions \eqref{Bndryv}-\eqref{term}.
We do not tackle directly the above equation \eqref{HJBv}, even if a characterization of $V$ as unique viscosity solution to it could be performed.\footnote{{One could try to prove the continuity of $V$, then show that $V$ is a viscosity solution of the HJB equation and finally use quite standard analytical techniques to prove a comparison in the viscosity sense for the equation and therefore get uniqueness for it (see e.g. \cite{CIL92, FS06, YZ}). Otherwise one could try to drop the proof of the continuity and deal with discontinuous viscosity solutions, for which the comparison is a bit harder to prove (see \cite[Ch.\,VII]{FS06}), and then prove the continuity a posteriori as a consequence of the characterization as viscosity solution. We will not do that,  since our study of the dual HJB equation will be sufficient to come back and prove a characterization of $V$ as classical solution to the HJB equation within a suitable class of smooth functions. Our uniqueness result will be weaker than what can be obtained by the viscosity approach, but will be enough for our purposes.}}  We just note here that this equation  is fully nonlinear and degenerate, so the regularity of its solutions cannot be obtained  dealing directly with it by the known methods of classical solutions of PDE's.\footnote{To this regard, we should mention, e.g.,  \cite{CTZ03, DFG11, Z94} for direct results in this direction, when the problem is autonomous and over an infinite horizon, and the equation elliptic. Up to our knowledge, despite a sketch in \cite{Z94}, there are no results of this kind for parabolic HJB equations coming from investment-consumption problems - as the one we deal with in this paper.}
What we can do is to apply duality to the problem and get a dual control problem with an associated  HJB equation for which we are able to prove regularity results.
For this purpose, consider, for $(t,y)\in [0,T)\times (0,+\infty)$, the sup-Legendre transform of ${U}^*_1(t,y,\cdot)$, i.e. the function (convex in $(y,u)$)
\beq\label{tf}
\widetilde{U_1^*}(t,y,{u})&:=&
\sup_{x\geq  0}\{{U}^*_1(t,y,x)- xu\}, \\
&=& \sup_{c,x \geq 0} \{U_1(t,c,x)-cy -  xu\}, \  \  \ (t,y,u)\in [0,T)\times (0,+\infty)\times(0,+\infty). \nonumber
\enq
For convenience of the reader, we notice that, when $U_1$ is separable in $x$ and $c$, i.e. $U_1(t,c,x)=U_1^{(1)}(t,c)\,+\,U_1^{(2)}(t,x)$, we have
\beqs
\widetilde{U_1^*}(t,y,{u}) &=& \widetilde{U_1}^{(1)}(t,y)\,+\,\widetilde{U_1}^{(2)}(t,u),
\enqs
where  $\widetilde{U_1}^{(1)},  \widetilde{U_1}^{(2)}$ are, respectively, the sup-Legendre transform of  ${U_1}^{(1)}, {U_1}^{(2)}$ with respect to the second variable.
Finally, we consider also the sup-Legendre transform of $U_2$, i.e. the function
\beq\label{tildeU2}
\widetilde{U_2}(y)&= &\sup_{x\geq 0} \left\{U_2(x)-xy\right\}, \ \ \ y>0.
\enq
Given $(t,y)\in\overline{D_T}$, we may consider a new
 control problem - which we call \emph{dual control problem}  and  denote  by \textbf{(D)} - that we are going to define (for the derivation of the argument see \cite[Sec.\,1]{R}).
Let $\beta=(\beta_s)_{s\in[t,T)}$ be a fixed adapted process with locally bounded integrable trajectories and
consider the controlled process $Y^{t,y,\beta,u}$ defined by the SDE
 \begin{equation}\label{eqqY}
 \begin{cases}
\displaystyle{dY_s\ =\   -u_s ds+\beta_s Y_sdB_s,}\\
Y_t\ =\ y,
\end{cases}
\end{equation}
with $u\;\in\;\mathcal{U}_\beta(t,y)$, where
\begin{multline}\label{tildeU}
\mathcal{U}_\beta(t,y)\ =\ \big\{({u}_s)_{s\in[t,T]} \ \mbox{is} \ (\mathcal{F}_s)_{s\in[t,T]}-\mbox{predictable, nonnegative, with integrable trajectories},\\ \mbox{ and such that} \ Y_s^{t,y,\beta,{u}}> 0 \ \mbox{a.s.} \ \forall s\in[t,T] \big\}.
\end{multline}
Let $x\in\R_+$, $y>0$ $(c,\pi)\in \mathcal{A}(t,x)$, $u\in\mathcal{U}_\beta(t, y)$, and set $X=X^{t,x,c,\pi}$ and $Y=Y^{t,y,\beta, u}$. Integration by parts yields
\beqs
d(X_s Y_s) &=& (-u_s X_s - c_sY_s) ds +Y_s \big( \pi_s \sigma(s) +\beta_sX_s) dB_s+ Y_s\pi_s(b(s)+\beta_s\sigma(s))ds.
\enqs
If
\beq\label{const}
b(s)+\beta_s\sigma(s)&=&0, \ \ \ \ \forall s\in[t,T],
\enq
 it follows that the process  $(X_s Y_s + \int_t^s (u_rX_r  + c_rY_r) dr)_{s \in[t, T]}$ is a supermartingale (as a positive local martingale), and in particular
\beq\label{pp}
\E \left[X_T Y_T + \int_t^T (u_s X_s +c_s Y_s) ds\right] &\leq& xy.
\enq
Now, by definition of $\widetilde{U_1^*}$ and $\widetilde{U_2}$  and by \eqref{pp}, if $Y_s>0$ almost surely for each $s\in[t,T]$, then
\beq\label{ppp}
&&\E \left[ \int_t^T U_1(s,c_s,X_s) ds + U_2(X_T)\right]\nonumber
\\ &\leq& \E \left[ \int_t^T \big(\widetilde{U_1^*}(s,Y_s,u_s) + c_s Y_s  +u_s X_s\big) ds + \widetilde{U_2}(Y_T) + X_T Y_T \right] \\
&\leq& \E \left[ \int_t^T \widetilde{U_1^*}(s,Y_s,u_s) ds + \widetilde{U_2}(Y_T)\right] + xy\nonumber .
\enq
Since  $(c,\pi)\in\mathcal{A}(t,x)$ is arbitrary, taking the supremum over  $(c,\pi)\in\mathcal{A}(t,x)$ on the left handside in \eqref{ppp}, we
get for every  $u\in\mathcal{U}(t,y)$
\beq
\label{pa}
V(t,x)&\leq& \E \left[ \int_t^T \widetilde{U_1^*}(s,Y_s,u_s) ds +\widetilde{U_2}(Y_T)\right] + xy.
\enq
Therefore, when \eqref{const} holds, the right handside of \eqref{pa} is an upper bound for the primal value function. On the other hand we can take  the infimum  over  $u\in\mathcal{U}_\beta(t,y)$ in the right handside of \eqref{pa}.
Taking into account that \eqref{pa} has been derived under \eqref{const},
this leads to consider the control problem
\beqs
\mbox{\textbf{(D)}} \ \ \ \ \ \  \inf_{u\in \mathcal{U}(t,y)}\widetilde{J}(t,y;u),
\enqs
where $\mathcal{U}(t,y)$ is the set defined in \eqref{tildeU} when $\beta$ is given by \eqref{const},
\beq\label{Jphi}
 \widetilde{J}(t,y;u)& =& \mathbb{E}\left[\int_t^{T}\widetilde{U_1^*}(s,Y_s^{t,y,u},{u}_s)ds+\widetilde{U_2}(Y_T^{t,y,u})\right],
\enq
and $Y^{t,y,u}$ is the solution to \eqref{eqqY} when $\beta$ is given by \eqref{const}, i.e. the solution to
\begin{equation}\label{eq:statedualpsi}
\begin{cases}
\displaystyle{dY_s\ =\  -{u}_sds-\frac{b(s)}{\sigma(s)}Y_sdB_s,}\\
Y_t\ =\ y.
\end{cases}
\end{equation}

We denote by $W$ the value function associated to this problem - that we call \emph{dual value function} - i.e.
\beq\label{ineqDual}
W(t,y)&:= & \inf_{{u}\in\mathcal{U}(t,y)}\widetilde{J}(t,y;u), \ \ \ \ \ \ \ (t,y)\in [0,T]\times(0,+\infty).
\enq
Taking the infimum over  $u\in\mathcal{U}(t,y)$ in the right handside of \eqref{pa} we get the inequality
\beq\label{ineqDual1} V(t,x)& \leq & W(t,y) + xy, \ \ \ \ \forall (t,y)\in [0,T]\times(0,+\infty).
 \enq
Defining the Legendre transform of the primal value  function
$$\widetilde{V}(t,y)\ \ :=\ \ \sup_{x\geq  0} \,\{V(t,x)-xy\}, \ \ \ (t,y)\in  [0,T]\times(0,+\infty),$$
from \eqref{ineqDual1} we get
\beq\label{eq}
\widetilde{V}&\leq & W, \ \ \ \ \mbox{on} \ \ { [0,T]\times(0,+\infty)}.
\enq
What one can  expect is the equality
\beq\label{eq2}
\widetilde{V}&=& W, \ \ \ \ \mbox{on} \ \ { [0,T]\times(0,+\infty)}.
\enq
We will prove \eqref{eq2} as corollary of our next results.

By standard stochastic control arguments we  associate to $W$ an HJB equation that we call dual HJB equation. It is the semilinear equation
\beq \label{eqDualHJB0}
-w_t(t,y)- \frac{b^2(t)}{2\sigma^2(t)}y^2 w_{yy}(t,y) -\inf_{u \geq 0} \widetilde{H}_{cv}(t,y,-w_y(t,y))&=&0,
\enq
where
\beq\label{ham2}
\widetilde{H}_{cv}(t,y,q)\ \ :=\ \  \widetilde{U^*_1}(t,y,u) + uq, \ \ \ q\in\mathbb{R}.
\enq
with terminal condition $w(T,\cdot)=\widetilde{U_2}.$
Since $U_1^*(t,y,\cdot)$ is concave over $\mathbb{R}_+$,
we have
{$$
U_1^*(t,y,x)\ =\ \inf_{u\geq 0} \,\{ \widetilde{U^*_1}(t,y,u)+ux\}, \ \ \ x>0.
$$
So, in the set where $w_y<0$ - it will be for every $(t,y)\in [0,T)\times (0,+\infty)$ in the case of our solution -  the HJB equation \eqref{eqDualHJB0} can be rewritten as
\beq \label{eqDualHJB}
-w_t(t,y)-\frac{b^2(t)}{2\sigma^2(t)}y^2 w_{yy}(t,y)-{U}^*_1(t,y,-w_y(t,y))&=&0.
\enq}

\begin{Remark}\label{rem:dual}
Due to the presence of current cost in the state (i.e. the dependence of $U_1$ on $x$), we have a (real) dependence of $\widetilde{U_1^*}$ on $u_s$ in the functional \eqref{Jphi} defining the dual problem. Since this dependence is  monotone (nonincreasing)
and
since $\widetilde{U_1^*}$ is also nonincreasing on $Y_s$ and $u_s$ appears with the negative sign in \eqref{eq:statedualpsi}, this creates a trade-off between the functional \eqref{Jphi} and the state equation \eqref{eq:statedualpsi}, giving rise to a real (nontrivial) control problem. At the level of the dual HJB equation \eqref{eqDualHJB0} above,  this can be appreciated by the presence of a nonlinearity in the first order term. When, as in \cite{BMZ, R}, the function $U_1$ does not depend on $x$,\footnote{Actually in \cite{BMZ} the function $U_1$ expressing the current utility is not even considered. However, as outlined in \cite{BMZ}, considering  a current utility depending \emph{only} on consumption would not complicate the mathematical problem.} the dependence of this term on $w_y$ disappears and the dual HJB equation is linear. While in \cite{BMZ} the linearity of the dual equation allows to deal with analytical solutions expressed through the heat kernel, a different and more theoretical approach is needed here. We are not aware of papers where the dual problem is investigated when also  utility on the current wealth is considered; nevertheless, we stress that  utility on the  current wealth arises in concrete problems, as the ones described in Section \ref{sec:app}.
\end{Remark}

\section{The dual value function as classical solution of the dual HJB equation}\label{sec:dual}

In this section we show that  $W$ is a classical solution to the HJB equation \eqref{eqDualHJB}. To do that first we show that it is a viscosity solution to  \eqref{eqDualHJB} and then we show its regularity.
\subsection{$W$ as viscosity solution of the dual HJB equation}
Before proceeding further, we need to investigate some properties of $\widetilde{U_1^*}$, $\widetilde{U_2}$ and derive qualitative properties for $W$.
\begin{Proposition}\label{prop:tildef}
We have the following properties of the functions  $\widetilde{U_1^*}$ and $\widetilde{U_2}$.
 \begin{enumerate}
 \item
  $\widetilde{U_1^*}:\R_+\times (0,+\infty)\times(0,+\infty) \rightarrow \R$  is nonnegative,  convex in $(y,u)$ and nonincreasing in $y$ and $u$.
\item
$\widetilde{U_2}: (0,+\infty) \rightarrow \R$    is nonnegative, convex and nonincreasing.
 \item We have the following growth estimate: there exists $\tilde{K}>0$ such that
\beq\label{growthU*}
\widetilde{U_1^*}(t,y,u) +\widetilde{U_2}(y) \ \ \leq \ \ \tilde{K} (1+y^{-\frac{p}{1-p}}+u^{-\frac{p}{1-p}}), \ \ \ t\in[0,T), \ u>0,\ y>0.
\enq
 \item We have
 \beq\label{tilde11}
(i)\ \ \lim_{{y\wedge u}\rightarrow {+\infty}}\widetilde{U_1^*}(t,y,u)\ =\ 0; \ \ \ \ \ \ \ (ii)\  \lim_{y\rightarrow{+\infty}}\widetilde{U_2}(y)\ =\ 0.
 \enq
{\item According to (a) or (b) of Assumption \ref{ass:U}(iv), we have respectively either
$$ (a) \ \ \ \ \exists\, \varepsilon >0 \ \mbox{such that} \ \lim_{y\rightarrow 0^+} \widetilde{U^*_1}(t,y,u)\ =\ +\infty \ \ \mbox{uniformly in } (t,u)\in [T-\varepsilon,T)\times \R_+,$$
or
$$ (b)  \ \ \ \ \ \lim_{y\rightarrow0^+}\widetilde{U_2}(y)\ =\ +\infty,$$
or both.
}

\item $U_1^*\in C_{loc}^{\delta/2, k + \delta}([0,T) \times (0,+\infty)\times (0,+\infty);\R)$, where $k\geq 2$ is the integer constant of Assumption \ref{ass:U}\,(i).
\end{enumerate}
\end{Proposition}
\textbf{Proof.} 1-2-3 follow straightly by using the properties of Legendre transforms and Assumption \ref{ass:U}(i,\,ii,\,iii).

\smallskip
4. For fixed $t>0$, let for $y>0, \ u>0$,
\beqs
\Lambda_{y,u} \ = \ \left\{ (x,c) \in \R_+^2 \ \  \Big| \ \ \frac{\partial}{\partial c} U_1(t,c,x) \;\geq\; y, \ \ \ \frac{\partial}{\partial x} U_1(t,c,x) \;\geq\; u\right\} \ \cup\  \{(0,0)\}.
\enqs
Using Assumption \ref{ass:U}(i), it is not difficult to see that the maximizer in the definition of $\widetilde{U_1^*}(t,y,u)$ belongs to $\Lambda_{y,u}$ and that $\Lambda_{y,u}$ shrinks to $\{(0,0)\}$ as $y\rightarrow +\infty$ and $u\rightarrow +\infty$; so
\beqs
\limsup_{{y\wedge u} \to +\infty}\  \widetilde{U_1^*}(t,y,u)\  \leq\  \limsup_{{y\wedge u} \to +\infty}\  \sup_{(c,x) \in \Lambda_{y,u}} U_1(t,c,x)\  =\  U_1(t,0,0) \ =\  0.
\enqs
The limit for $\widetilde{U_2}$ follows with a similar argument.

\smallskip
{5. If we are in the case of Assumption \ref{ass:U}(iv)(a), then, due to monotonicity with respect to $u$ of $\widetilde{U_1^*}$,  the statement (a) is equivalent to
\beq\label{ihg}
\exists \varepsilon> 0 \ \ \ \mbox{such that} \ \ \ \lim_{y\rightarrow 0} \lim_{u\rightarrow+\infty} \widetilde{U^*_1}(t,y,u)=+\infty, \ \ \mbox{uniformly w.r.t.} \ t\in[T-\varepsilon,T).
\enq
Now, by 
  \eqref{growthU},
   using the same argument of point 4 above, but with respect to $u$ only, we get
\beq\label{vxc}
\lim_{u\rightarrow+\infty} \widetilde{U^*_1}(t,y,u)\ \ = \ \ \sup_{c\geq 0} \,\{U_1(t,c,0)-cy\},  \ \mbox{uniformly w.r.t.} \ t\in[T-\varepsilon,T).
\enq
Since taking $c=1/y$ we get
\beq\label{pwx}
\sup_{c\geq 0}\{U_1(t,c,0)-cy\} & \geq &
U_1(t,1/y,0)-1,
\enq
 the claim \eqref{ihg} follows combining \eqref{vxc}-\eqref{pwx} and using Assumption \ref{ass:U}(iv)(a).

In the case of Assumption \ref{ass:U}(iv)(b) the claim (b) can be obtained  as above (but more easily) by using the definition \eqref{tildeU2}.

\smallskip
6. If Assumption \ref{ass:U}(i)(b) holds, the claim is immediate as
$$
U_1^*\ =\ U_1\ \in \ C_{loc}^{\delta/2, k + \delta}([0,T) \times (0,+\infty)\times (0,+\infty);\R).
$$
 Let us prove the claim in the case when  Assumption \eqref{ass:U}(i)(a) holds true.
 Under our assumptions, the map  $c\mapsto \frac{\partial}{\partial c} U_1(t,\cdot,x)$ is a bijection from $(0,+\infty)$ to $(0,+\infty)$  for each $(t,x)$ $\in$ $[0,T) \times (0,+\infty)$, and the supremum in the definition of $U_1^*$ is attained at the unique $c^*(t,y,x)$ satisfying
\beq \label{defc^*}
\frac{\partial}{\partial c}\  U_1(t,c^*(t,y,x),x)\  \ =\ \  y.
\enq
Since $\frac{\partial}{\partial{c^2}}U_1$ $<$ $0$, it follows from the implicit function theorem that $c^*$ has the same regularity properties as $\frac{\partial}{\partial c} U_1$, i.e. it is $C^{\delta/2, k -1 + \delta}_{loc}([0,T) \times (0,+\infty)\times(0,+\infty);\R)$.
Writing $$U_1^*(t,y,x) \ \ =\ \  U_1(t,c^*(t,y,x),x) - c^*(t,y,x) y$$ and using \reff{defc^*}, we obtain
\beqs
\frac{\partial}{\partial y} \ U_1^*(t,y,x) &=& -c^*(t,y,x),\\
\frac{\partial}{\partial x} \ U_1^*(t,y,x) &=& \frac{\partial}{\partial x} U_1(t,c^*(t,y,x),x).
\enqs
Both of these functions lie  in $C_{loc}^{\delta/2, k -1 + \delta}([0,T) \times (0,+\infty)\times(0,+\infty);\R)$, which proves the claim.
\hfill$\square$

\begin{Proposition}\label{prop:W}
$W$ is finite, strictly positive on $D_T$, convex and strictly decreasing in $y$. Moreover, we  have the growth condition, for some $K_W>0$,
\beq\label{gw}
W(t,y) &\leq &K_W ( 1 + y^{- \frac{p}{1-p}}), \ \ \ \ \forall (t,y)\in [0,T]\times (0,{+\infty}),
\enq
and terminal and boundary conditions
\beq\label{btw}
\begin{cases}
(i) \ \ W(T,y) \ =\ \widetilde{U_2}(y), \ \ \ \ \ \ \ \ \ \ \ \, \ \ \forall y\in(0,+\infty);\\ (ii)
\ \lim_{y \to 0^+} \ W(t,y) \ =\   {+\infty}, \ \ \   \ \,\forall t\in[0,T);\\
(iii) \ \lim_{y \to  {+\infty}} \ W(t,y) \ =\  0, \ \ \ \ \  \ \forall t\in[0,T].
\end{cases}
\enq
\end{Proposition}

%
%
\textbf{Sketch of proof.} The arguments are quite standard and we only sketch the proof of the claims which are straightforward.

Taking the feedback control $u_s = Y_s$ in the state equation \eqref{eq:statedualpsi} and using \eqref{growthU*}, we obtain that $W$ is finite and satisfies the  growth condition \eqref{gw}.
The
strict positivity in $D_T$ is more tricky and we give a complete proof, which follows from Proposition \ref{prop:tildef}(5). Indeed, let $(t,y)\in D_T$. Since $Y^{t,y,u}_T\leq Y^{t,y,0}_T$ for each $u\in\mathcal{U}(t,y)$, we get
\beq\label{okj}
\widetilde{J}(t,y;u)\ \ \geq\  \ \E\left[\int_{t}^T\widetilde{U_1^*}(s,Y^{t,y,0}_s,u_s)ds+\widetilde{U_2}(Y_T^{t,y,0})\right], \ \ \  \forall u\in  \mathcal{U}(t,y).
\enq
Since $Y^{t,y,0} $ is a Geometric Brownian Motion, setting
$$
A^{t,y}_{\varepsilon,y_0}\ \ :=\ \ \Big\{\sup_{s\in [{t\vee(T-\varepsilon)}, T]}Y^{t,y,0}_s<y_0\Big\},
$$
we have
\beq
\label{prob}
p_{\varepsilon,y_0}^{t,y}\ :=\ \P(A^{t,y}_{\varepsilon,y_0}) \ > 0, \ \ \ \ \forall \varepsilon>0, \ \forall y_0>0.
\enq
Now,
set for all $(s,y_0)\in [0,T)\times (0,+\infty)$
\beq\label{ggs}
g(s,y_0)\ \ :=\ \ \lim_{u\rightarrow+\infty} \widetilde{U_1^*}(s,y_0,u).
\enq
Using \eqref{okj}, \eqref{prob} and \eqref{ggs}, we get
\beq\label{okj1}
\widetilde{J}(t,y;u)\ \ \geq\ \ p_{\varepsilon,y_0}^{t,y} \left[\int_{{t\vee(T-\varepsilon)}}^Tg(s,y_0)ds+\widetilde{U_2}(y_0)\right], \ \ \  \forall u\in  \mathcal{U}(t,y).
\enq
Now, if Assumption \ref{ass:U}(iv)(a) holds, take $\varepsilon$ above as the one in appearing in the same assumption. By  Proposition \ref{prop:tildef}(5)(a), we can choose $y_0>0$ such that $g(s,y_0)\geq \delta$ for all $s\in[t,T]$ for a suitable $\delta >0$. Since \eqref{okj1} is uniform in $u\in  \mathcal{U}(t,y)$, we get the claim in this case. If we assume that Assumption \ref{ass:U}(iv)(b) holds, then  from it,  \eqref{okj1} and Proposition \ref{prop:tildef}(5)(b) still follows the claim.

Convexity comes from convexity of $\widetilde{U_1^*}$ and $\widetilde{U_2}$, and from  linearity of the state equation by standards arguments. Also monotonicity is consequence of standard arguments due to monotonicity of $\widetilde{U_1^*}$ and $\widetilde{U_2}$.

The terminal condition \eqref{btw}(i) comes from the definition of $W$ immediately.

{The boundary condition  \eqref{btw}(ii) can be  obtained arguing as in the proof of strict positivity of $W$. Indeed,  we can consider \eqref{okj1} with $y_0=y$.
Then, since $Y^{t,y,0}=yY^{t,1,0}$ we get that
\beq	\label{plls}
p_{\varepsilon,y}^{t,y} \ = \ p_{\varepsilon,1}^{t,1}\ >\ 0, \ \ \  \forall y\in(0,1).
\enq
 Therefore, \eqref{okj1}
 becomes in this case
 \beq\label{okj2}
\widetilde{J}(t,y;u)\ \ \geq\ \ p_{\varepsilon,1}^{t,1} \left[\int_{{t\vee(T-\varepsilon)}}^Tg(s,y)ds+\widetilde{U_2}(y)\right], \ \ \  \forall u\in  \mathcal{U}(t,y).
\enq
from which we get
 \beq\label{okj3}
W(t,y)\ \ \geq\ \  p_{\varepsilon,1}^{t,1}\,\left[ \int_{{t\vee(T-\varepsilon)}}^Tg(s,y)ds+\widetilde{U_2}(y)\right].
\enq
Taking the limit for $y\rightarrow 0^+$ and using  Proposition \ref{prop:tildef}(5), we get  \eqref{btw}(ii). }

Let us show now the boundary condition \eqref{btw}(iii). Let $(t,y)\in [0,T]\times (0,+\infty)$ and take the feedback control  $u_s =  Y_s$ in \eqref{eq:statedualpsi} and consider the associated state trajectory $Y_s^{t,y,u}$. Then
\beq\label{Wop}
W(t,y)&\leq &\widetilde{J}(t,y; u)\ \ = \ \ \E\left[\int_{t}^T\widetilde{U_1^*}(s,Y^{t,y,u}_s,Y^{t,y,u}_s)ds+\widetilde{U_2}(Y_T^{t,y,u})\right].
\enq
Since
$$
Y_s^{t,y,u}\ \ =\ \ y \cdot \mbox{exp}\left(-\int_t^s \big(1+\frac{b^2(\xi)}{2\sigma^2(\xi)}\big)d\xi-\int_t^s \frac{b(\xi)}{\sigma(\xi)}dB_\xi\right),
$$
we have
\beq\label{expp}
Y_s^{t,y,u} &\rightarrow & +\infty, \ \ \  \forall s\in[t,T], \ \mbox{a.s.}
\enq
Hence,
using \eqref{tilde11} and \eqref{expp}
we get
\beq\label{limlim}
\widetilde{U_1^*}(s,Y^{t,y,u}_s,Y^{t,y,u}_s) \  \rightarrow \  0, \ \  \forall s\in[t,T], \ \mbox{a.s.}, \ \ \mbox{and} \ \ \ \ \widetilde{U_2}(Y_T^{t,y,u}) \ \rightarrow  \  0, \ \ \mbox{a.s.}
\enq
On the other hand, thanks to {\eqref{growthU*}}, we have
\beqs\label{domdom}
\widetilde{U_1^*}(s,Y^{t,y,u}_s,Y^{t,y,u}_s)  \ \leq  \  \tilde{K}\left(1+2(Y_s^{t,y,u})^{-\frac{p}{1-p}}\right),\ \ \ \
 \widetilde{U_2}(Y_T^{t,y,u})\ \leq \ \tilde{K}(1+(Y_T^{t,y,u})^{-\frac{p}{1-p}}).
\enqs
Since the above right hand sides are integrable uniformly in $y\geq 1$, using \eqref{Wop}  and \eqref{limlim} we get the claim by Vitali's Theorem.

Finally, strict monotonicity follows from convexity, monotonicity, strict positivity and   \eqref{btw}(iii).
\hfill$\square$


%
%
%


\begin{Proposition} \label{propContW}
$W$ is continuous on $[0,T]\times (0,{+\infty})$. Moreover $W(\cdot,y)$ is nondecreasing for all $y\in(0,+\infty)$.
\end{Proposition}

\textbf{Proof.}
First of all, by convexity, $W$ is continuous in the space variable $y$ for each fixed $t\in[0,T]$.

Let us show continuity in time. For that, we need to exploit the following Dynamic Programming Principle:\footnote{Appealing to the Dynamic Programming Principle may seem somehow unfair, as usually it is problematic to prove it if one has not proved before the continuity of the value function (and we are just proving the continuity invoking it). However, we observe that in this case (where the time $t'$ is deterministic)  the proof of the Dynamic Programming Principle (see, e.g., \cite[Ch.\,4]{YZ}), only uses the continuity in the space variable $y$.}
for each $t,t'$ such that $0\leq t\leq t'\leq T$ and each $y$ $\in$ $(0,+\infty)$,
\beq \label{eqDPPW}
W(t,y) &=& \inf_{u \in \mathcal{U}(t,y)} \E \left[\int_t^{t'} \widetilde{U_1^*}(s,Y^{t,y,u}_s,u_s)ds +W(t',Y^{t,y,u}_{t'})\right].
\enq
Now we show that $W$ is nonincreasing in time.
 Indeed, let $(t,y)\in[0,T)\times (0,+\infty)$, let $u\in\mathcal{U}(t,y)$  and let $t'\in[t,T]$. Since $\widetilde{U_1^*}$ $\geq$ $0$, from \eqref{eqDPPW} we have
\beq\label{11}
W(t,y)
& \geq &
\inf_{u \in \mathcal{U}(t,y)} \E \left[W(t',Y^{t,y,u}_{t'})\right].\enq
By monotonicity of $W$ in $y$ and since $Y_{t'}^{t,y,u}\leq Y_{t'}^{t,y,0}$ for all $u\in\mathcal{U}(t,y)$, we get
\beq\label{12}
\inf_{u \in \mathcal{U}(t,y)} \E \left[W(t',Y^{t,y,u}_{t'})\right]  &\geq&  \E \left[ W(t', Y_{t'}^{t,y,0}) \right].
\enq
Combining \eqref{11} and \eqref{12}, and using Jensen's inequality, we finally get
$$W(t,y)\ \ \geq \ \ W(t',y),$$
proving the monotonicity claim.

From this monotonicity it follows that the functions
provided by the left and right limits of $W$ in $t$, i.e.
$$W_+(t, \cdot) \ :=  \ \lim_{h \downarrow 0} W(t+h,\cdot), \ \ \ \ \ \  \ W_-(t, \cdot) \ :=  \ \lim_{h \downarrow 0} W(t-h,\cdot),$$
are well-defined in $[0,T)$ and $(0,T]$ respectively, and
\beq\label{ineq0}
W_-\ \ \geq\ \  W\ \ \geq \ \ W_+
\enq
(where the functions are defined).
We note that $W_+,W_-$ are also convex in $y$ for fixed $t$, so they are  continuous in $y$ for fixed $t$ as well.
If we show the inequalities
\beq\label{ineq1}
W_-\ \ \leq\ \  W\ \ \leq \ \ W_+
\enq
(where the functions are defined)
combining with \eqref{ineq0} the proof of continuity in time will be complete.

Let us first show the left inequality in \eqref{ineq1}. For any $s\in[0,T]$, define $\widehat Y^{s,y}$ as the process corresponding to the feedback control $\hat u_\cdot = \widehat Y_\cdot$ starting from $(s,y)$. Then,  for each $r\geq s$,
\beqs
\widehat Y_r^{s,y} &=&  y \exp \left( \int_s^r (-1 + \frac{1}{2} \frac{b(\xi)^2}{\sigma(\xi)^2}) d\xi - \int_s^r \frac{b(\xi)}{\sigma(\xi)} dB_\xi  \right).
\enqs
Note that, since $\frac{b(\cdot)}{\sigma(\cdot)}$ is bounded, we have the following estimates\,:
\beq
\E \left[ \left|\widehat Y_r^{s,y} - y \right| \right] &\leq& \omega(|s-r|), \ \ \mbox{ with } \omega \ \mbox{continuous and }  \omega(0^+)=0, \label{ineqY1}\\
\sup_{0 \leq r \leq s \leq T} \E \left[ |\widehat Y_r^{s,y}|^{q}\right] &<& + \infty, \;\;\; \ \forall q \in \R. \label{ineqY2}
\enq
Let $t\in[0,T]$ and take a sequence $t_n \uparrow t$. By \reff{eqDPPW} and \eqref{growthU*},
\beq
W(t_n,y) &\leq& \E \left[ \int_{t_n}^t \widetilde{U_1^*}(s,\widehat Y^{t_n,y}_s,\widehat Y^{t_n,y}_s)ds + W(t,\widehat Y^{t_n,y}_{t})\right]  \nonumber \\
&\leq& \E\left[ \int_{t_n}^t 2\tilde{K}(1+ (\widehat Y^{t_n,y}_s)^{- \frac{p}{1-p}})ds + W(t,\widehat Y^{t_n,y}_{t})\right]. \label{ineqWt-}
\enq
By \reff{ineqY2} the expectation of the integral in \eqref{ineqWt-} goes to $0$. On the other hand, from \reff{ineqY1}, passing to a subsequence if necessary (we have monotonicity in $t$, so we can do that without loss of generality), we see that $\widehat Y^{t_n,y}_t$ $\to$ $y$ almost surely. Hence, using \reff{ineqY2} and the growth condition \reff{gw} on $W$, by dominated convergence we get
\beqs
\lim_{n \to \infty} \E \left[ W(t,\widehat Y^{t_n,y}_{t})\right] &=& W(t,y).
\enqs
So, we finally obtain $W_-(t,y) \leq W(t,y)$.

Now let us turn to the proof of the right inequality in \eqref{ineq1}. Let $t\in[0,T)$ and take a sequence $t_n \downarrow t$.  Again, using \reff{eqDPPW} we have that
\beqs
W(t,y) &\leq&\E\left[\int_{t}^{t_n} \widetilde{U_1^*}(s,\widehat Y^{t,y}_s,\widehat Y^{t,y}_s)ds + W(t_n,\widehat Y^{t,y}_{t_n})\right].
\enqs
The proof is now the same once we  show that $W(t_n, \widehat Y^{t,y}_{t_n})$ $\to$ $W_+(t,y)$ almost surely. We observe that  $W(t_n, \cdot)$ $\searrow$ $W_+(t,\cdot)$ pointwise by definition. Since all these functions are continuous,
 by Dini's Theorem we get $W(t_n, \cdot)$ $\searrow$ $W_+(t,\cdot)$ locally uniformly.  Therefore $t_n \downarrow t$, $y_n$ $\to$ $y$ implies $W(t_n,y_n) \to W_+(t,y)$. Since, by passing to a subsequence if necessary (again we may do that without loss of generality because of monotonicity in $t$) we can assume $Y^{t,y}_{t_n}$ $\to$ $y$ almost surely, it follows that $W(t_n, \widehat Y^{t,y}_{t_n})$ $\to$ $W_+(t,y)$ almost surely. And again by dominated convergence this implies $W(t,y) \leq W_+(t,y)$. This completes the proof of continuity in time.

Now it just remains to notice that again by Dini's Theorem the continuity of $W$ in $t$ is locally uniform in $y$, which combined to the fact that $W$ is continuous in $y$ for fixed $t$, implies joint continuity of $W$ in $(t,y)$.
\hfill$\square$\\\\
Now we may state the viscosity property of $W$.
\begin{Proposition} \label{propDualVisc}
$W$ is a continuous viscosity solution to \reff{eqDualHJB} in $D_T$.
\end{Proposition}

{\bf Proof.} Due to continuity of $W$, this is quite standard. We omit the proof  for brevity and refer to classical references, such as \cite{FS06,YZ}.
\hfill$\square$
\subsection{Regularity of $W$} \label{sec:reg}
In this section we prove a regularity result for the   dual value function $W$.
\begin{Theorem}\label{prop:regpsic12}
\begin{enumerate}
 \item []
 \item
 $W\in C_{loc}^{1+\frac{\delta}{2},k+2+\delta}(D_T;\R)$.
 \item
 $W_y(t,\cdot)<0$,  \  ${W}_y(t,0^+)=-{\infty}$ and ${W}_y(t,{+\infty})=0$, for every $t\geq 0$.
 \item
 $W_{yy}>0$ over $D_T$.
\end{enumerate}
\end{Theorem}

\ni{\bf Proof.}
{\bf 1.}
Take any $(t_0,y_0)\in D_T  $  and consider, for suitable $\varepsilon >0$, the square
$$D_\varepsilon(t_0,y_0)\ :=\ [t_0,t_0+\varepsilon)\times (y_0-\varepsilon,y_0+\varepsilon)\
\subset\ D_T.$$
First of all, note that, due to convexity, the right and left space derivatives of $W$ exist. Denoting them by $W_y(t,y^+)$ and $W_y(t,y^-)$ respectively,  again by convexity we have $W_y(t,y^+)\geq W_y(t,y^-)$. Moreover,  there exist $M_\varepsilon$, $m_\varepsilon$ $>$ $0$ such that
\beq\label{loclip}
M_\eps \ \geq\ \sup_{(t,y) \in D_\varepsilon(t_0,y_0)} - W_y(t,y^-) \ \geq \ \inf_{(t,y) \in D_\varepsilon(t_0,y_0)} -W_y(t,y^+) \ \geq \ m_\eps . \label{ineqWy}
\enq
Indeed, by convexity  $-W_y(t,y^+) \geq \frac{1}{y}(W(t,y) - W(t,2y))$, and, since $W$ is continuous and strictly decreasing in $y$ for each $t$, the infimum above must be strictly positive. In the same way, $-W_y(t,y^-) \leq -\frac{2}{y}(W(t,y) - W(t,y/2))$ and the supremum is finite.

By Proposition \ref{propDualVisc}, the dual value function ${W}$ is a viscosity solution of the dual HJB equation
\eqref{eqDualHJB} in $D_\varepsilon (t_0,y_0)$ with Dirichlet boundary condition
\begin{equation}\label{eq:BCforHJBdualforreg}
{w}\ =\ {W}, \qquad \mbox{on} \quad {\cal P}(D_\varepsilon (t_0,y_0)),
\end{equation}
where ${\cal P}(D_\varepsilon (t_0,y_0))$ is the parabolic boundary of $D_\varepsilon (t_0,y_0)$ defined as
$$
{\cal P}(D_\varepsilon (t_0,y_0))\ :=\ \{t_0+\varepsilon\}\times [y_0-\varepsilon,y_0+\varepsilon]
\cup  [t_0,t_0+\varepsilon]\times  \{y_0-\varepsilon,y_0+\varepsilon \}.
$$
Consider the function $F$ defined on $D_\varepsilon (t_0,y_0) \times \R$ by
$$F(t,y,q)\ :=\ U_1^*\big(t,y, -[(m_\varepsilon \vee q)\wedge M_\varepsilon]\big).$$
By Proposition \ref{prop:tildef}(6),  $F$ is H\"older continuous in $D_\varepsilon (t_0,y_0) \times \R$.
By \reff{ineqWy}, we have that  $W$ is actually a viscosity solution in  $D_\varepsilon (t_0,y_0)$ to the equation
\beq \label{eqDualHJBLoc}
-w_t(t,y)-\frac{b^2(t)}{2\sigma^2(t)}y^2 w_{yy}(t,y)-F(t,y,w_y(t,y))&=&0.
\enq
Since ${W}$ is continuous on ${\cal P}(D_\varepsilon (t_0,y_0))$, then we have uniqueness of  viscosity solutions
to \eqref{eqDualHJBLoc} with boundary condition \eqref{eq:BCforHJBdualforreg}  (see, e.g., \cite[Cor.\,8.1,\,Ch.\,V]{FS06}). On the other hand, due to Assumption \ref{ass:b,sigma} and to H\"older continuity of $F$, the PDE \eqref{eqDualHJBLoc} is semilinear uniformly parabolic on $D_\varepsilon (t_0,y_0)$ with H\"older continuous coefficients, so  by
  Theorem 12.22  of \cite{Lieberman}  - with the assumptions of Theorem 12.16 of the same book - it admits a solution fulfilling the boundary condition \eqref{eq:BCforHJBdualforreg} in the space $C^{1,2}(D_\varepsilon(t_0,y_0);\R)$. This (classical) solution is also a viscosity solution, thus, due to uniqueness of viscosity solutions,  it coincides with $W$.  Hence, we conclude that $W\in C^{1,2}(D_\varepsilon(t_0,y_0);\R)$, therefore,
by  arbitrariness of $(t_0,y_0)$, that $W$ $\in$ $C^{1,2}(D_T;\R)$.

Given that, we know that $-W_y$ is strictly positive and locally Lipschitz continuous in $D_T$. Moreover, by Proposition \ref{prop:tildef}(6),  $U_1^*\in C_{loc}^{\delta/2, k + \delta}([0,T) \times (0,+\infty)\times(0,+\infty);\R)$. Therefore, the claim  follows from a simple induction, using  regularity results for linear equations of the form $-u_t - Lu=f$ (see, e.g., Theorem 8.12.1, p.\,131, in \cite{KryHolder}).

\smallskip
{\bf 2.} The first claim follows \eqref{loclip}. The other ones follow from convexity and from \eqref{btw}(ii) and \eqref{btw}(iii), respectively.

\smallskip

\textbf{3.}
As in \cite{BMZ} we use a maximum principle argument.
Differentiating twice \reff{eqDualHJB},  we get
\begin{multline*}
- (W_{yy})_t - \frac{b^2(s)}{2\sigma^2(s)}\left[2W_{yy}+4y(W_{yy})_y+y^2(W_{yy})_{yy}\right]\\ -(U_1^*)_{yy}(t,y,-W_y) + W_{yyy}\cdot (U_1^*)_{x}(t,y,-W_y)  \\+ 2 W_{yy} \cdot (U_1^*)_{xy}(t,y,-W_y) - W_{yy}^2 \cdot(U_1^*)_{xx}(t,y,-W_y)\  =\  0.
\end{multline*}
Noting that $U_1^*$ is convex in $y$, we see that $W_{yy}$ is a nonnegative supersolution to the linear parabolic PDE
\begin{multline*}
- u_t -  \frac{b^2(s)}{2\sigma^2(s)}\left[2u+4yu_y+y^2u_{yy}\right] + (U_1^*)_{x}(t,y,-W_y) u_y\\ + [2 (U_1^*)_{xy}(t,y,-W_y) - W_{yy}\cdot (U_1^*)_{xx}(t,y,-W_y)] u \ = \ 0.
\end{multline*}
Hence, by a strong maximum principle (see e.g. \cite[Th.\,3, Ch.\,II]{Fri}), if $W_{yy}(t_0,y_0)=0$ for some $(t_0,y_0)\in D_T$, it must be $W_{yy} \equiv 0$ on $(t_0,T) \times (0,{+\infty})$, which is clearly in contradiction, e.g.,  with \eqref{btw}(ii).
\hfill$\square$\\
\ni From Proposition \ref{propDualVisc} and Theorem \ref{prop:regpsic12} we get the following
\begin{Corollary}\label{Wregsol}
$W$ is a classical solution to \reff{eqDualHJB} in $D_T$.
\end{Corollary}
\section{Back to the primal control problem: verification and optimal controls}
 Let $t\in [0,T]$ and let $\widetilde{{W}}$ be the inf-Legendre transform of ${W}(t,\cdot)$, i.e.
\beq\label{ttildepsi}
\widetilde{{W}}(t,x)&:=&\inf_{y> 0}\,\{{W}(t,y)+xy\}, \ \ \ \ (t,x)\in \overline{D_T}.
\enq
Due to its definition and to the positivity of $W$ (see Proposition \ref{prop:W}), the function $\widetilde{{W}}$ is finite and nonnegative on $\overline{D_T}$.  Moreover, it is concave and nondecreasing in $x$ for each $t\in[0,T]$ and, due to Theorem \ref{prop:regpsic12}, it can be written, for $(t,x)\in D_T$,
as
\beq\label{PHI*}
\widetilde{{W}}(t,x)&=&{W}\left(t,[{W}_y(t,\cdot)]^{-1}(-x)\right)+x\,[{W}_y(t,\cdot)]^{-1}(-x).
\enq
We are going to prove that
\beq\label{eq21}
\widetilde{W} & =& V,\ \ \ \ \mbox{on} \ \overline{D_T}
\enq
(we notice that \eqref{eq21}  implies, as corollary,  \eqref{eq2}, i.e. $\widetilde{V}=W$) and that $V$ is the unique classical solution of the primal HJB equation  \eqref{HJBv} in the following class:
\begin{multline*}
\mathcal{C} \ \ =\ \ \Big\{ v\in C(\overline{D_T};\R)\cap C_{loc}^{1+\delta/2,k+2+\delta}(D_T;\R)\
 \mbox{such that }
  \ v_x>0, \ v_{xx}<0 \ \mbox{in} \ D_T, \\ \mbox{and} \  v \ \mbox{fulfills the boundary and growth conditions \eqref{bdrw} below}   \Big\}
\end{multline*}
where
\beq\label{bdrw}
\begin{cases}
(i) \ \ {v}(t,0) \ \ = \ \ 0, \ \ \ \ \ \forall t\in[0,T], \\
(ii) \ \  {v}(T,x)\ \ = \ \ U_2(x), \ \  \ \ \ \forall x\geq 0,\\
(iii)  \ \ \exists K_0\ \mbox{such that} \ 0\ \leq \  {v}(t,x)\  \leq\  K_0 (1+ x^p), \ \ \ \ \ \forall  (t,x) \in [0,T]\times[0,+\infty).
\end{cases}
\enq
We note that if  $v\in \mathcal{C}$, due to \eqref{hamiltspecial}, we have
 \beq\label{hamiltspecial2}
H(t,x,v_x(t,x),v_{xx}(t,x))  &=& {U}^*_1(t,v_x(t,x),x)- \frac{b^2(t)}{2\sigma^2(t)}\frac{v_x(t,x)^2}{v_{xx}(t,x)}.
\enq
 We proceed as follows:
\begin{enumerate}
\item We show that $\widetilde{W}\in\mathcal{C}$ and that it is a classical solution of the primal HJB equation \eqref{HJBv} (Proposition \ref{prop:tildepsi}).
\item We show that a verification theorem holds for \textbf{(P)} for every classical solution $v\in\mathcal{C}$ of the primal HJB equation (Theorem \ref{teo:ver}).
\item We show that  for every classical solution $v\in \mathcal{C}$ of the primal HJB equation the associated closed loop equation admits a solution and that this implies  $v=V$ (Proposition \ref{prop:CLE} and Corollary \ref{Vchar}).
\end{enumerate}
Clearly, these three points yield the equality $\widetilde{W}= V$ and the announced uniqueness.

\subsection{$\widetilde{W}$ as a classical solution of the primal HJB equation}
\begin{Proposition}\label{prop:tildepsi}
$\widetilde{W}\in\mathcal{C}$
and solves  the primal HJB equation \eqref{HJBv} in classical sense in $D_T$.
Moreover
%
 it satisfies the Inada conditions in $x$:
$$
\widetilde{W}_x(t,0^+)\ =\ +\infty, \  \ \ \ \widetilde{W}_x(t,+\infty)\ =\ 0, \ \ \ \ \ \forall t\in[0,T).
$$

\end{Proposition}
\textbf{Proof.}
\emph{Growth and boundary conditons.}
{The growth condition \eqref{bdrw}(iii) follows from \eqref{ttildepsi} and \eqref{gw}.
The boundary condition \eqref{bdrw}(i) follows from \eqref{ttildepsi} and \eqref{btw}(iii).
The boundary condition \eqref{bdrw}(ii) follows from  \eqref{ttildepsi},    \eqref{btw}(i) and the fact that the inf-Legendre transform of $\widetilde{U_2}$ is $U_2$.}

{
\emph{Continuity in $\overline{D_T}$.}
The fact that $\widetilde{{W}}$ is continuous in $D_T$  follows from \eqref{PHI*} and Theorem \ref{prop:regpsic12}.
Now we show the continuity at the boundary $[0,T)\times \{0\} $.

Continuity of $\widetilde{W}(t,\cdot)$ at $0^+$ for each $t\in[0,T)$ follows from \eqref{ttildepsi}: it yields
\beqs
\widetilde{W}(t,x) & \leq & W(t,\varepsilon/x)+\varepsilon, \ \ \ \forall x\geq 0, \ \forall \varepsilon>0,
\enqs
hence, taking into account also that $W$ is nonnegative and \eqref{btw}(iii),
$$0\  \leq \ \limsup_{x\downarrow 0}\widetilde{W}(t,x)  \ \leq  \  \varepsilon,  \ \ \ \forall \varepsilon>0,
$$
and, since $\varepsilon$ is arbitrary and taking into account  \eqref{bdrw}(i), we may conclude that
$$\lim_{x\downarrow0} \widetilde{W}(t,x) \ = \ 0\  = \ \widetilde{W}(t,0).$$ Moreover, by monotonicity of $\widetilde{W}(t,\cdot)$ for all $t\in[0,T)$ the convergence above is locally uniform in $t\in[0,T)$ due to Dini's Theorem, so, combining with the obvious continuity of $\widetilde{W}(\cdot,0)$, we get the continuity of $\widetilde{W}$ at the boundary $[0,T)\times \{0\}$ in the couple $(t,x)$.

Next we show the
continuity at the boundary $\{T\}\times\R_+$. First let us show the continuity of $W(\cdot,x)$ at $T^-$ for fixed $x\in\R_+$.  Since $\widetilde{W}(t,0)=0$ for every $t\in[0,T]$, the claim is obvious for $x=0$, so we now assume $x$ $>$ $0$.
Clearly,  for any $y>0$,
$$\limsup_{t \uparrow T}\ \widetilde  W(t,x) \ \leq\ \limsup_{t \uparrow T}\ \{W(t,y) + xy \} \ =\  W(T,y)+xy,$$
 by continuity of $W$. Taking the infimum over $y$, we obtain the inequality
 $$\limsup_{t \uparrow T} \ \widetilde W(t,x)\  \ \leq\  \ \widetilde{W}(T,x).$$
 For the opposite inequality, we notice that, by definition of $W$, we have for each $y > 0$ and each $t\in[0,T]$
$$W(t,y) \ \geq \  \E \left[ \widetilde{U_2}(Y_T^{t,y,0})\right] \ \geq \  \widetilde{U_2}\left(\E\left[ Y_T^{t,y,0}\right]\right) \ =\ \widetilde{U_2}(y),$$
where we have used Jensen's inequality. Since $\widetilde{W}(T,\cdot)=U_2(\cdot)$, we get $W(t,\cdot)\geq W(T,\cdot)$, which in turn yields
$$\liminf_{t \uparrow T} \ \widetilde W(t,x)\ \  \ge \ \ \widetilde{W}(T,x).$$ Now, taking into account the obvious continuity of $\widetilde{W}(T,\cdot)$ in $\mathbb{R}_+$, the continuity of $\widetilde{W}$ at the boundary $\{T\}\times \R_+$ in the couple $(t,x)$ follows again from Dini's Theorem, as $\widetilde{W}(\cdot,x)$ inherits from $W(\cdot,y)$ the monotonicity (Proposition \ref{propContW}).  This concludes the proof of the continuity of $\widetilde{W}$ on $\overline{D_T}$.}

{
\emph{Further regularity in $D_T$.} From \eqref{PHI*} and taking into account Theorem \ref{prop:regpsic12}, we get
for each $(t,y)\in D_T$
\beq\label{www}
\begin{cases}
(i) \ \ \widetilde{W}_t(t,x)\ =\ W_t\left(t,[W_y(t,\cdot)]^{-1}(-x)\right),\\\\
(ii)
\ \ \widetilde{W}_x(t,x)\ =\ [W_y(t,\cdot)]^{-1}(-x),\\\\
(iii) \ \ \displaystyle{\widetilde{W}_{xx}(t,x)\ =\ -\frac{1}{W_{yy}(t,[W_y(t,\cdot)]^{-1}(-x))}.}
\end{cases}
\enq
So, due to Theorem  \ref{prop:regpsic12}, we have
 $\widetilde{{W}}\in C_{loc}^{1+\delta/2,k+2+\delta}(D_T;\R)$ and   $\widetilde{{W}}_x>0 , \widetilde{{W}}_{xx}<0$ in $D_T$. This completes the proof that $\widetilde{{W}}\in\mathcal{C}$. }

{\emph{$\widetilde{W}$ as solution to the HJB equation}. The fact that $\widetilde{{W}}$ solves the HJB equation \eqref{HJBv} in classical sense in $D_T$ follows from  Corollary \ref{Wregsol}  by straightforward computations using \eqref{hamiltspecial2} and \eqref{www}.}


{\emph{Inada's conditions.} Inada's conditions follow from Theorem \ref{prop:regpsic12}(2) and  \eqref{www}(ii).}
\hfill$\square$
\subsection{Verification theorem}\label{sub:ver}
\begin{Theorem}\label{teo:ver}
Let $v\in \mathcal{C}$ be a classical solution to the primal HJB equation \eqref{HJBv}.
Then:
\begin{itemize}
\item[(i)] $v(t,x)\geq V(t,x)$ for all $(t,x)\in\overline{D_T}$.
\item[(ii)] Let $(t,x)\in\overline{D_T}$,   let $(c^*,\pi^*)\in\mathcal{A}(t,x)$ and let $X^*:=X^{t,x,c^*,\pi^*}$. If
\beq\label{HVH}
H_{cv}(s,X^*_s,v_x(s,X^*_s),v_{xx}(s,X^*_s);\, c^*_s,\pi^*_s) & =& H(s,X^*_s,v_x(s,X^*_s),v_{xx}(s,X^*_s))
\enq
$\P$-almost surely for almost every  $s\in [t,T]$,
then $(c^*,\pi^*)$ is an optimal control and $v(t,x)=V(t,x)$.
\end{itemize}
\end{Theorem}

{\bf Proof.}
(i)
Let $(t,x)$ $\in$ $\overline{D_T}$, $(c,\pi)$ $\in$ $\Ac(t,x)$, and,  to simplify the notation, let us  write $X_s:=X^{t,x,c,\pi}_s$ for all $s\in[t,T]$. Set
$$\tau\ \ := \ \ \inf\ \{ s \in [t,T]  \ | \ X_s=0 \} \wedge T.$$
We notice that, due to the state constraint, $\mathcal{A}(s,0)=\{(0,0)\}$ for all $s\in[t,T]$ and the corresponding state trajectory is identically $0$, so
\beq\label{csd}
\mbox{if} \ \ \tau<T,\ \  \mbox{then} \ (c, \pi,X)\ \equiv \ (0,0,0)\ \ \mbox{in the random time interval}  \ [\tau,T].
\enq
Now we may find a sequence of stopping times $\tau_n \nearrow \tau$ such that $\int_0^\cdot v_x(s,X_s) \pi_s \sigma(s) dB_s$ is a martingale in $[t,\tau_n]$.
Since $v\in  C^{1,2}([t,T) \times (0,+\infty);\mathbb{R})$ and satisfies the HJB equation \reff{HJBv},  It\^o's formula yields
\beqs
\E\left[ v(\tau_n, X_{\tau_n}) \right] &=& v(t,x) \ +\  \E\bigg[ \int_t^{\tau_n} \big(H_{cv}-H)(s,X_s,v_x(s,X_s),v_{xx}(s,X_s); c_s,\pi_s)ds\bigg] \nonumber \\
&& \;\;\;\;\;\;\;\;\;\;\ \, - \  \E\,\bigg[\int_t^{\tau_n} U_1(s,c_s,X_s)\big) ds \bigg] \nonumber \\
&\leq&  v(t,x) - \E\left[ \int_t^{\tau_n} U_1(s,c_s,X_s) ds \right].
\enqs
This gives us
\beq
v(t,x)\ \  \geq \ \ \E\left[ v(\tau_n, X_{\tau_n}) + \int_t^{\tau_n} U_1(s,c_s,X_s) ds \right], \ \  \ \ \forall n\in\mathbb{N}. \label{eqVer}
\enq
Letting $n$ $\to$ $\infty$ in \eqref{eqVer}, using Fatou's Lemma on the first term of the expectation of the right handside, and monotone convergence on the second one, we  get
\beq\label{exx}
v(t,x) &\geq& \E\left[ v(\tau,X_\tau) + \int_t^\tau U_1(s,c_s,X_s) ds \right]\nonumber\\
&=& \E\left[\mathbf{1}_{\{\tau<T\}} \Big(v(\tau,X_\tau) + \int_t^\tau U_1(s,c_s,X_s) ds \Big)\right]\\
&&+\, \E\left[\mathbf{1}_{\{\tau=T\}} \Big(v(\tau,X_\tau) + \int_t^\tau U_1(s,c_s,X_s) ds \Big)\right].\nonumber
\enq
Using \eqref{csd}, the fact that $U_2(0)=0$ and  that $U_1(\cdot,0,0)=v(\cdot,0)=0$, we get
\beq\label{exx2}
v(t,x) &\geq&  \E\left[ U_2(X_T) + \int_t^T U_1(s,c_s,X_s) ds \right].
\enq
Since $(c,\pi)\in\mathcal{A}(t,x)$ was arbitrary, this means that $v(t,x) \geq V(t,x)$, and (i) is proved.
\smallskip

(ii) Let $(c^*,\pi^*)\in\mathcal{A}(t,x)$ satisfying  \eqref{HVH}, and denote $X^*=X^{t,y,c^*,\pi^*}$. In this case we have equality in \eqref{eqVer}, i.e.
\beq\label{pds}
v(t,x) &=&   \E\left[ v(\tau_n, X^*_{\tau_n}) + \int_t^{\tau_n} U_1(s,c^*_s, X^*_s)ds \right], \ \  \ \ \ \forall n\in\mathbb{N}.
\enq
Now we take the limit for $n\rightarrow \infty$ keeping the equality above. We cannot use Fatou's Lemma as before for the part $v(\tau_n,X_{\tau_n})$, but we need to use a result keeping the equality in the limit. Since $\lim_{n \to \infty} v(\tau_n, X^*_{\tau_n}) = v(\tau,X^*_\tau)$ almost surely,  it suffices to prove uniform integrability of $(v(\tau_n,X^*_{\tau_n}))_{n\geq 0}$.
For this purpose,
write $Y_s$ $:=$ $Y_s^{t,1,0}$ for all $s\in[t,T]$. We know from the discussion following \reff{const} that  $\big(X_s^*Y_s + \int_t^s c^*_u Y_u du\big)_{s\in[t,T]}$ is a supermartingale.  Since $c_s^* Y_s \geq 0$, we see that also  $(X_s^*Y_s)_{s\in[0,T]}$ is a supermartingale,  hence  $\E[X^*_{\tau_n} Y_{\tau_n}]\leq x$. Now, taking $q\in(p,1)$, we get, using \eqref{bdrw}(iii) ,
\beqs
\E \left[v(\tau_n,X^*_{\tau_n})^{q/p}\right] &\leq& \E\left[K_0^{q/p} (1+|X^*_{\tau_n}|^p)^{q/p}\right]
\ \ \leq \ \    K_0^{q/p} 2^{\frac{q}{p}-1}\big(1+\E[|X^*_{\tau_n}|^q]\big).
\enqs
Now, using  H\"older's inequality, from the inequality above we get
\beqs
\E \left[v(\tau_n,X^*_{\tau_n})^{q/p}\right]
&\leq& K_0\big(1+ \E[X^*_{\tau_n} Y_{\tau_n}]^q \E [ (Y_{\tau_n})^{- \frac{q}{1-q}}]^{1-q} \big) \ \ \leq\ \  K'_0\big(1+ x^{q}\big).
\enqs
So the sequence $v(\tau_n, X^*_{\tau_n})_{n\geq 0}$ is bounded in $L^{q/p}$ with $q/p>1$. By de La Vall\'ee Poussin's Theorem it is uniformly integrable. Hence taking the limit in \eqref{pds} we get
\beq\label{pds1}
v(t,x) &=&   \E\left[ v(\tau, X^*_{\tau}) + \int_t^{\tau} U_1(s,c^*_s, X^*_s)ds \right].
\enq
Splitting on the sets $\{\tau<T\}$ and $\{\tau=T\}$ as above, taking into account that $v(T,\cdot)=U_2(\cdot)$ for the part corresponding to set  $\{\tau=T\}$,  taking into account \eqref{csd} and  that $v(\cdot,0)=0=U_1(\cdot,0,0)$ on the set $\{\tau<T\}$, we finally rewrite \eqref{pds1} as
\beq\label{VJ}
v(t,y)&=& J(t,y;c^*,\pi^*).
\enq
Combining  \eqref{VJ} with the claim (i)  we get the claim (ii).
\hfill$\square$\\\\
From Proposition \ref{prop:tildepsi} and Theorem \ref{teo:ver}, we see that $\widetilde{W}\geq V$.\footnote{This inequality may be also proved using  \eqref{eq} and the concavity of $V$ in $x$ which could be proved directly.}
What we want to get is indeed the equality, and in order to get it we need to exploit further item (ii) of Theorem \ref{teo:ver} finding optimal feedback controls.

\subsection{Optimal feedback controls}\label{sub:clos}
Given  $v\in\mathcal{C}$,  we may define feedback maps in classical sense associated to the maximization of
$H_{cv}$ in the HJB equation \eqref{HJBv}.  They are, for $s\in [0,T)$,
\beq\label{C*}
{C}^{v}(s,x)\ =\begin{cases}
\ \begin{cases} \left[\frac{\partial}{\partial c}U_1(t,\cdot,x)\right]^{-1}(v_x(t,x)), \ \ \, \mbox{if} \ x>0,\\ 0, \ \ \ \ \ \ \  \ \ \ \ \ \ \ \ \ \qquad\qquad\qquad\mbox{if} \ x=0,
\end{cases}   \mbox{if Assumption \ref{ass:U}(i)(a) holds,}\\\\
0, \ \ \ \  \ \ \ \ \ \mbox{if Assumption \ref{ass:U}(i)(b) holds,}
\end{cases}
\enq
\beq\label{Pi*}
\Pi ^{v}(s,x)&=&\begin{cases}-\frac{b(s)v_x(s,x)}{\sigma(s)v_{xx}(s,x)}, \ \ \ \mbox{if} \ x>0,\\
0, \ \ \ \ \ \ \ \ \ \ \ \ \ \ \ \ \   \mbox{if} \ x=0.
\end{cases}
\enq
Their definition for $x>0$ is indeed given by the maximization of $H_{cv}$ in the HJB equation taking into account the structure of the Hamiltonian \eqref{hamiltspecial} for functions in $\mathcal{C}$, while the definition at $x=0$ is due to the  the state constraint, which implies ${\mathcal{A}}(t,0)=\{(0,0)\}$. \\\\
The closed loop  equation associated to the feedback maps $C^{v},\Pi^{v}$ is
\beq\label{CLE}
\begin{cases}
dX_s= -{C}^v(s,X_s)ds+b(s)\Pi^v(s,X_s)ds+ \sigma(s)\Pi^v(s,X_s)dB_s,\\
X_t=x.
\end{cases}
\enq
Since $v\in\mathcal{C}$, one has local Lipschitz continuity of $\Pi^v(s,\cdot)$ on $(0,{+\infty})$ for every $s\in[t,T)$.
and  local Lipschitz continuity of $C^v(s,\cdot)$ on $(0,{+\infty})$ for every $s\in[t,T)$.
{We notice that, since we have defined the coefficients $\Pi^v(s,\cdot)$and  $C^v(s,\cdot)$ only on $\mathbb{R}_+$, we only look for nonnegative solutions to the above equations.}
\begin{Proposition}\label{prop:CLE}
Given $v\in\mathcal{C}$ and $(t,x)\in[0,T)\times \R_+$, there exists a unique (nonnegative) solution $X^{t,x;v}$ to the closed loop equation
\eqref{CLE} in the interval $[t,T]$.
\end{Proposition}
\textbf{Proof.}
\emph{Existence.} If $x=0$ the claim is clear, just by taking  $X^{t,x;v}\equiv 0$. Let $x>0$. Due to local Lipschitz continuity of $C^v(s,\cdot), \Pi^v(s,\cdot)$, using standard SDE's theory (see, e.g., \cite[Ch.\,5,\,Th.\,2.9 ]{KS-bm}), we get for each $\varepsilon>0$  the existence of a unique solution $X^{t,x,\varepsilon;v}\in [\varepsilon,\varepsilon^{-1}]$ in the stochastic  interval $[t,\tau_\varepsilon)$, where $\tau_\varepsilon$ is implicitly defined in terms of the solution itself as
\beqs
\tau_\varepsilon&=&\inf\,\{s\in[t,T] \ | \ X_s^{t,x,\varepsilon;v}\leq \varepsilon \mbox{ or } X_s^{t,x,\varepsilon;v}\geq \varepsilon^{-1}\},
\enqs
with the convention $\inf \emptyset =T$.
Of course, if $\varepsilon<\varepsilon'$,  we have $\tau_\varepsilon>\tau_{\varepsilon'}$ and
\beq\label{pppp}
 X_s^{t,x,\varepsilon}&\equiv& X_s^{t,x,\varepsilon'} \ \ \mbox{on}\ \  [t,\tau_{\varepsilon'}), \ \ \  \ \ \forall \ 0<\varepsilon<\varepsilon'.
\enq
Set $$\tau\;=\;\lim_{\varepsilon\downarrow 0}\tau_\varepsilon.$$
Then by \eqref{pppp} there exists a unique solution  $X^{t,x,v}\geq 0$ to \eqref{CLE} in the interval $[t,\tau)$.
We now show that this solution can be extended to the whole interval $[t,T]$.
By a Girsanov transformation (note that the Novikov condition holds true due to our assumptions on $b,\sigma$), there exists a probability $\Q$  equivalent to $\P$, and a $\Q$-Brownian motion $\tilde{W}$, such that  \reff{CLE} may be rewritten as
\beqs
dX_s\;=\; -{C}^v(s,X_s)ds+ \sigma(s){\Pi}^v(s,X_s)d\tilde{W}_s.
\enqs
By nonnegativity of $C^v$, the process $X^{t,x;v}$ is a nonnegative $\Q$-supermartingale on $[t,\tau)$, which can be extended to a $\Q$-supermartingale  {($L^1$ bounded)}  on $[t,T]$  setting it equal $0$ in $[\tau,T]$. Hence, by Doob's convergence Theorem (see e.g. \cite[Theorem II.2.5]{RY})
, there exists a finite random variable $X^{t,x;v}_{\tau}$ such that
\beq \label{ineqQZeps1}
\lim_{s \nearrow \tau} X^{t,x;v}_s \ = \ X^{t,x;v}_{\tau}, \;\;\;\;\;\;\;\ \Q\mbox{-a.s.}.
\enq
Since $\Q$ $\sim$ $\P$, we also have
\beq \label{ineqQZeps}
\lim_{s \nearrow \tau} X^{t,x;v}_s \ = \ X^{t,x;v}_{\tau}, \;\;\;\;\;\;\;\ \P\mbox{-a.s.}.
\enq
Immediately  \eqref{ineqQZeps} yields the desired extension on $\{\tau=T\}$. Let us now consider the set $\{\tau<T\}$.
On this set we have $X^{t,x;v}_{\tau_\eps}$ $\in$ $\{\eps, \eps^{-1}\}$, so that by \eqref{ineqQZeps} necessarily $X^{t,x;v}_{\tau} = 0$ almost surely, getting
\beq \label{limZtau}
\lim_{s \nearrow \tau} X^{t,x;v}_s \;= \;0 \;\;\;\;\; \mbox{a.s. on } \ \{\tau<T\}.
\enq
Therefore, we may now extend $X^{t,x;v}$ to a solution defined over $[t,T]$ on $\{\tau<T\}$ by setting
$$X_s^{t,x;v}\ \equiv\ 0, \ \ \mbox{for}\  s\in [\tau,T].$$

\emph{Uniqueness.}  Let $Y^{t,x;v}\geq 0$ be another solution in  $[t,T]$. First, in view of the proof of the existence part, we have $Y^{t,x;v}=X^{t,x;v}$ in $[t,\tau]$, where $\tau$ is the random time defined in the existence part. Moreover, since $X^{t,x;v}_\tau=0$, we  also  have $Y^{t,x;v}_\tau=0$. Then, since $Y^{t,x;v}$  is a nonnegative $\Q$-supermartingale as solution of \eqref{CLE}, it must be $Y^{t,x;v}\equiv 0$ in $[\tau,T]$, concluding the proof (as also $X^{t,x;v}\ \equiv\ 0$ in $[\tau,T]$).
\hfill$\square$
{\begin{Remark}
Notice that in the proof of Proposition \ref{prop:CLE} we strongly use two  facts:
\begin{enumerate}
\item the coefficients $C^v(t,\cdot), \Pi^v(t,\cdot)$ are defined only on $\mathbb{R}_+$, hence we look for solutions only in the class of nonnegative processes;
 \item  the coefficient $C^v(t,\cdot)$ is nonnegative, hence the solution (under $\mathbb{Q}$) is a supermartingale.
\end{enumerate}
Also we notice that we do not need the continuity of the maps $C^v(t,\cdot), \Pi^v(t,\cdot)$ at $0^+$.
\end{Remark}}
\begin{Corollary}\label{Vchar}
We have $\widetilde{W}=V$ and it is the unique  solution in $\mathcal{C}$ to the HJB equation \eqref{HJBv}. Moreover, given $(t,x)\in[0,T]\times \mathbb{R}_+$, an optimal control in feedback form for \textbf{(P)} starting at $(t,x)$ is given by
\beq\label{optcon}
c_s^*\ \ =\ \ C^{V}(s,X^{t,x;V}_s), \ \ \ \ \ \ \pi_s^* \ \ =\ \ \Pi^{V}(s,X^{t,x;{V}}_s),
\enq
where $C^V, \Pi^V$ are the feedback maps defined in \eqref{C*}-\eqref{Pi*} associated to $V\in\mathcal{C}$, and where $X^{t,x;{V}}$ is the unique solution to \eqref{CLE} associated to $C^V, \Pi^V$.
\end{Corollary}
{
\textbf{Proof.}
By Proposition \ref{prop:tildepsi}, we know that $\widetilde{W}\in\mathcal{C}$ and solves  the HJB equation \eqref{HJBv}. On the other hand given any solution $v\in\mathcal{C}$ to \eqref{HJBv}, for any given $(t,x)\in[0,T)\times \mathbb{R}_+$ we can construct by Proposition \ref{prop:CLE} a solution $X^{t,x;v}\geq 0$ to the closed loop equation \eqref{CLE}. Defining the feedback controls
\beqs
c_s^*\ \ =\ \ C^{v}(s,X^{t,x;v}_s), \ \ \ \ \ \ \pi_s^* \ \ =\ \ \Pi^{v}(s,X^{t,x;{v}}_s),
\enqs
by uniqueness we have  $X^*:= X^{t,x; c^*,\pi^*} = X^{t,x;v}$ and the triple $(X^*,c^*,\pi^*)$ satisfies by construction \eqref{HVH}. Then applying   Theorem \ref{teo:ver} we conclude $v=V$.
\hfill$\square$}
\begin{Remark}
As consequence of Proposition \ref{prop:tildepsi} and Corollary \ref{Vchar}, we see that  $V$ satisfies the Inada condition $\frac{\partial}{\partial x}\,V(t,0^+)=+\infty$ even if $\frac{\partial U_1}{\partial c}(\cdot,0^+)$, $\frac{\partial U_1}{\partial x}(\cdot,0^+)$ and $U_2'(0^+)$ (which are well defined by concavity) are all finite.  Indeed,  the fact that $V$ satisfies the Inada condition at $0^+$ is simply due to the fact that $x=0$ is an absorbing boundary combined with Assumption \ref{ass:U}(iv).
\end{Remark}


%
%
%
%
\subsection{An alternative way to optimality : probabilistic duality}
In the previous parts of the current section we have constructed the  optimal control couple \eqref{optcon} by exploiting the duality at an analytical level to study the regularity of the primal value function $V$. This approach seems particularly meaningful from a PDE point of view, as it produces a regularity result for the degenerate fully nonlinear PDE \eqref{HJBv}.

However, to construct optimal controls for the primal problem $\textbf{(P)}$ it is not strictly needed to study the regularity of $V$, as they can be obtained starting from the construction of optimal controls for the dual control problem $\textbf{(D)}$ and then exploiting further the duality argument
of Section \ref{sec:primdual} that led to the definition of the dual control problem $\textbf{(D)}$.

We illustrate in this subsection this alternative (probabilistic) dual way to optimality\footnote{{The authors are indebted to one anonymous Referee who suggested this alternative approach.}}, which is based on the following steps.

\begin{enumerate}
\item One constructs, by Dynamic Programming arguments, an optimal feedback control $u^*$ for the dual control problem $\textbf{(D)}$.
\item Considering the optimal state/control couple $(Y^*,u^*)$ for  $\textbf{(D)}$, one tries to define a control/state triple $(X^*,c^*, \pi^*)$ for $\textbf{(P)}$ such that, plugging  $(Y^*,u^*)$ and $(X^*,c^*, \pi^*)$, the inequalities in \eqref{ppp} become equalities.
\item Finally, one deduces the optimality of the triple  $(X^*,c^*, \pi^*)$ for the primal control problem $\textbf{(P)}$.
\end{enumerate}

\textbf{Step 1.}
Consider the feedback map associated to the minimization  of \eqref{ham2}, i.e.
(cf. Theorem \ref{prop:regpsic12} for the well-posedness of this definition and notice that $G$ is nonnegative)
$$G(t,y): =\  \mbox{argmin}_{u\geq 0}\left\{ \widetilde{U_1^*}(s,y,u)- uW_y(s,y)\right\}, \ \ \ (t,y)\in[0,T)\times (0,+\infty).$$
i.e.
$$G(t,y)\: =\  \frac{\partial}{\partial x} U_1^*(s,y,-W_y(s,y)), \ \ \ (t,y)\in[0,T)\times (0,+\infty).
$$

The following result can be proved  using arguments similar to the ones used in Subsections \ref{sub:ver} and \ref{sub:clos}. We do not prove it for the sake of brevity,  limiting ourselves to  few remarks after the statement. 

\begin{Theorem}\label{teo:sub}
Let $(t,y)\in[0,T)\times (0,+\infty)$.
\begin{enumerate}
\item
The closed loop state equation associated to $G$
\beq\label{dualCLE}
\begin{cases}
\displaystyle{dY_s\ =\  -{G}(s,Y_s) ds-\frac{b(s)}{\sigma(s)}Y_sdB_s,}\\
Y_t\ =\ y,
\end{cases}
\enq
admits a unique solution $Y^{t,y;G}>0$ over $[t,T]$.
\item  The feedback control
\beq \label{ocd}
u_s^*:= G(s,Y_s^{t,y;G}), \ \ \ s\in[t,T],
\enq
belongs to $\mathcal{U}(t,y)$ and is optimal for the dual control problem $\textbf{(D)}$ starting from $(t,y)$.
\end{enumerate}
 \end{Theorem}
 \begin{Remark}\label{remteo}
\begin{itemize}
\item[]
\item[(i)] We do not really have  to prove a verification theorem for $W$, as we already know that $W$ is a classical solution to the dual HJB equation \reff{eqDualHJB} (cf. Corollary \ref{Wregsol}); this means that the analogue of the part (i) of the proof of Theorem \ref{teo:ver} does not need to be proved for all the admissible controls but only for the candidate optimal ones;
\item[(ii)] Since the control problem consists in minimizing positive quantities, the passage to the limit of a localizing sequence can be done with Fatou's Lemma and does not require any uniform integrability.
\item[(iii)] Let us detail a bit the proof of  of Theorem \ref{teo:sub}. The existence and uniqueness of a nonnegative solution $Y^{t,y;G}$ can follow the line of the proof of Proposition \ref{prop:CLE} once one shows the local Lipschitz continuity with respect to  $y$ in $(0,+\infty)$ and extending $G$ for $y=0$ by setting it equal to $0$. Instead, to prove the strict positivity one can follow two paths.
\begin{itemize}
\item[(a)] Studying
the behavior of this map at $y=0^+$. For example, if one is able to prove that this map is sublinear in a right neighborhood of   $0$, then one can compare the solution with a stochastic exponential and then get its strict positivity.
\item[(b)] Using martingale arguments as follows.
Define, with the convention $\inf\emptyset=T$,
$$
\tau\ :=\ \inf\{s\in[t,T] \ | \ Y^{t,y;G}_s=0\}.
$$
By applying It\^o's formula, using the fact that $W$ solves the HJB equation  \reff{eqDualHJB}  and the fact that $Y^{t,y;G}$ solves the closed loop equation \reff{dualCLE}, one gets as usual in verification arguments  that
$$\left(W(s,Y^{t,y;G}_s) + \int_t^s  \widetilde{U_1^*}(r,Y^{t,y;G}_r,u^*_r)dr\right)_{s\in[t,\tau)}$$ is a local martingale. Since it is nonnegative and since the integrand above is also nonnegative, it follows that $\left(W(s,Y^{t,y;G}_s)\right)_{s\in[t,\tau)}$   is a supermartingale. The latter implies $\lim_{s\rightarrow \tau^-} W(s,Y^{t,y;G}_s)<\infty$ almost surely. Due to \eqref{btw}(ii) and monotonicity of $W(s,\cdot)$, this is equivalent to  $\lim_{s\rightarrow \tau^-}Y^{t,y;G}_s>0$, and then we conclude  $Y^{t,y;G}>0$ over $[t,T]$.
\end{itemize}
\end{itemize}
 \end{Remark}
 \textbf{Step 2.}
 Let $(t,y)\in[0,T)\times (0,+\infty)$, consider the optimal control $u^*$ for  $\textbf{(D)}$ starting from $(t,y)$  defined in \eqref{ocd} and the associated state process $Y^{*}:=Y^{t,y,u^*}=Y^{t,y;G}$.
Considering the first inequality of \eqref{ppp} and plugging into it the couple $(Y^*,u^*)$, in order to get optimality  for the primal problem, we need to  fill the duality gap. To this aim,  we need first of all to choose, if possible, an admissible triple $(X^*,c^*,\pi^*)$ - where $X^*= X^{t,x,c^*,\pi^*}$ - such that this inequality becomes an equality when plugging $(X^*,c^*)$ into it, i.e.

 \beq\label{ppp0}
&&\E \left[ \int_t^T (U_1(s,c^*_s,X^*_s)-  c^*_s Y^*_s  -u^*_s X^*_s) ds + U_2(X^*_T)-  X^*_T Y^*_T\right]\nonumber
\\ & =& \E \left[ \int_t^T \widetilde{U_1^*}(s,Y^*_s,u_s^*)  ds + \widetilde{U_2}(Y^*_T) \right].
\enq
This is done by defining the process
\beq \label{kjhg}
X_s^*&:=&-W_y(s,Y^*_s), \ \ \ s\in[0,T).
\enq
Using Theorem  \ref{prop:regpsic12} and Corollary \ref{Wregsol}, the differentiation with respect to $y$ of \reff{eqDualHJB} and an application of It\^o's formula to \eqref{kjhg} yield $X^* = X^{t,x,c^*,\pi^*}$, where
$$
x:= X^*_t=-W_y(t,y), \ \ \ \  c^*_s\ :=\ -\frac{\partial}{\partial y}\widetilde{U_1^*}(s,Y_s^*,u_s^*), \ \ \
\pi_s^*\ :=\  \frac{b(s)}{\sigma^2(s)} Y^*_sW_{yy}(s,Y^*_s).
$$
Noting that, by definition of $u^*$, \eqref{kjhg} is equivalent to $X^*_s = - \frac{\partial}{\partial u}\widetilde{U_1^*}(s,Y_s^*,u_s^*)$,  we see that
\beq\label{kkjh}
U_1(s,c^*_s,X^*_s)-  c^*_s Y^*_s  -u^*_s X^*_s &=& \widetilde{U_1^*}(s,Y^*_s,u_s^*), \;\; \ \P\otimes ds -\mbox{ a.e. in } \Omega\times [0,T);
\enq
In addition \eqref{kjhg} is also equivalent to
\beq
\label{eq:dualXY}
W(s,Y^*_s) + X^*_s Y^*_s &=& \widetilde{W}(s,X^*_s) ,  \;\; \;\  \P\otimes ds- \mbox{a.e. in }\;\; \Omega\times [0,T).
\enq
Letting $s \to T$ in \eqref{eq:dualXY}, we conclude, by \eqref{btw}(i), concavity of $U_2$ - which ensures that the inf-Legendre transform of $\widetilde{U_2}$ coincides with $U_2$ -  and continuity of $X^*_\cdot Y^*_\cdot$, that
$$\widetilde{U_2}(Y^*_T) \ =\  U_2(X^*_T) - X^*_T Y^*_T, \ \ \ \mbox{a.s..}$$  Hence, by \eqref{kkjh} and \eqref{eq:dualXY}, the equality  \eqref{ppp0} is proved.

Now note that, by \eqref{eq:dualXY},  \eqref{ttildepsi} and \eqref{gw},  one has
$$X^*_s Y^*_s\ \leq \ \widetilde{W}(s,X^*_s) \ \leq\ K(1+ |X^*_s|^p),  \;\; \;\;\;\; \;\; \;\  \P\otimes ds- \mbox{a.e. in }\;\; \Omega\times [0,T).$$ Then, we can use the same argument as in the proof of Theorem \ref{teo:ver}\,(ii) to show that $\left(X^*_s Y^*_s + \int_t^s (u^*_r X^*_r +c^*_r Y^*_r) dr\right)_{t\leq r \leq T}$ is in fact a uniformly integrable martingale, so that \eqref{pp} holds with equality in this case, i.e.
  \beq\label{pp0}
\E \left[X^*_T Y^*_T + \int_t^T (u^*_s X^*_s +c^*_s Y^*_s) ds\right] &=& xy.
\enq
Then, combining \eqref{ppp0} and \eqref{pp0},  we deduce
\beq\label{ppp1}
\E \left[ \int_t^T U_1(s,c^*_s,X^*_s) ds + U_2(X^*_T)\right]
& =& \E \left[ \int_t^T \widetilde{U_1^*}(s,Y^*_s,u^*_s) ds + \widetilde{U_2}(Y^*_T)\right] + xy.
\enq

\textbf{Step 3.}
Using the  optimality of $(Y^*,u^*)$ and \eqref{ineqDual1}, from \eqref{ppp1} we get
\beqs\label{ppp2}
\E \left[ \int_t^T U_1(s,c^*_s,X^*_s) ds + U_2(X^*_T)\right]
& =& W(t,y) + xy\ \geq \ V(t,x),
\enqs
providing the optimality of $(c^*,\pi^*)$.
\section{Applications}\label{sec:app}
Current utility on the wealth may arise in several situations. For instance, we mention pension funds allocation (see, in a context of utility maximization,  \cite{DFG11,F11} and, in a context of quadratic cost minimization, \cite{DFGV,GHV}); optimal portfolio problems with random horizon (see \cite{BEJM, BP04}); markets with illiquidity (see \cite{FG, FGG}). We are going to describe the latter two applications.
%
\subsection{Portfolio optimization with random horizon} \label{subsecrandom}
A first application of our framework is to portfolio problems with random horizon. Consider the consumption/investment problem with state equation \eqref{wealth} when the time horizon of the agent is $T\wedge \tau$ where $T>0$ is fixed and  $\tau$ is some random variable $\tau\in[0,+\infty)$, i.e. the objective to maximize is a functional such as
\beq \label{funrandom}
\E\left[\int_0^{\tau\wedge T} G_1(t,c_t)dt+ G_2(\tau\wedge T,X_{\tau\wedge T})\right].
\enq
In this context it  is meaningful to assume, in general, that $\mathcal{F}_T \neq\mathcal{F}$, and  that $\tau$ is just  $\mathcal{F}$-measurable. A special case, which is the one we illustrate, as it may be covered by our framework, is when $\tau$ is independent of $\mathcal{F}_T$ (this problem has been already treated in \cite{BEJM} in the case of terminal utility).
 Since $\tau$ is independent of $(\mathcal{F}_t)_{t\geq 0}$,
setting $F(t)=\P\,\{\tau\leq t\}$ and assuming that $F$  admits a density $f$ over $[0,T)$,  the functional \eqref{funrandom} may be rewritten as\footnote{See \cite{EJY} for the rewriting of the term corresponding to $G_2$ in the general case when $\tau$ may be dependent on  $\mathcal{F}_{T}$, in which case one has to consider $F(t) :=\P\,\{\tau\leq t\ |\ \mathcal{F}_t\}$.}
 \beq \label{funrandom2}
\E\left[\int_0^T (G_1(t,c_t)(1-F(t))+G_2(t,X_t)f(t))dt+(1-F(T))G_2(T,X_T)\right].
\enq
So, it falls into our setting \ -\   under suitable assumptions on the functions $G_1,G_2$ \ -\  with
\beqs
U_1(t,c,x)&=& G_1(t,c)(1-F(t))+G_2(t,x)f(t), \\
U_2(x)&=& (1-F(T))G_2(T,x).
\enqs
Therefore we can apply our  results, which allow to construct optimal feedback controls by  Corollary \ref{Vchar}. To this regard we notice that in \cite{BEJM} the regularity of the value function is \emph{assumed} in the verification theorem, so the results given through the Dynamic Programming approach in \cite{BEJM} are definitively based on the possibility of finding  (regular) explicit solutions to the HJB equation. Hence, while in \cite{BEJM} it is needed to take specific structures for the utility function, here we do not need that.

Finally, we observe that the rewriting of \eqref{funrandom} as \eqref{funrandom2} can be performed also in the case $T=\infty$. So, applying our  Remark \ref{rem:negp}(iv), we get that our  results on the HJB equation and on the optimal feedback controls hold also in this case. The next subsection provides a significant example.
\subsection{Investment/consumption problems in markets with  illiquid assets}

A related application of our results is the mixed liquid/illiquid investment model studied in \cite{FG,FGG}. We refer to the latter references for details on the model.

Consider a market constituted by a riskless asset (assumed constant), and two risky assets $L$ and $I$ following Black-Scholes dynamics:
\beqs
dL_t & = & L_t\,( b_Ldt + \sigma_LdW_t),  \ \ \ \ L_0\ =\ 1,\\
dI_t & = & I_t\,\Big(b_I dt + \sigma_I\, (\rho dW_t + \sqrt{1 - \rho^2} dB_t)\Big),\ \ \ \ I_0\ =\ 1,
\enqs
where $W$ and $B$ are independent Brownian motions, and $\rho \in (-1,1)$ is a correlation parameter.

The specificity of the model is that, while the liquid asset $L$ may be observed and traded continuously, the illiquid asset $I$ may only be traded and observed at discrete random times $(\tau_k)_{k\geq 0}$, where we assume that $\tau_0=0$, and the interarrival times $\tau_{k+1} - \tau_k$ are i.i.d., and independent from $(B,W)$.

The investor's strategy is then a triple $((c_t)_{t\geq 0},(\pi_t)_{t\geq 0},(\alpha_k)_{k\in\mathbb{N}})$ where the components represent, respectively, the consumption, the amount invested in the liquid asset $L$ at time $t$, and the amount invested in the illiquid asset  $I$ at time $\tau_k$. The investor's wealth then follows the dynamics
\beqs
R_0 &=& r, \label{stateR1}\\
R_t &=& R_{\tau_k} + \int_{\tau_k}^{t} \big(\pi_s (b_L ds + \sigma_L dW_s)- c_s ds\big)  + \alpha_k \left(\frac{I_{t}}{I_{\tau_k}} -1\right), \ \ t\in(\tau_k,\tau_{k+1}].\label{stateR2}
\enqs
The investor  aims at optimizing the following criterion
\beqs
V(r) &=& \sup_{(c_t,\pi_t,\alpha_k)\in \mathcal{A}(r)} \E \int_0^\infty e^{- \beta s} U(c_s) ds,
\enqs
where $U$ is a  utility function, the discount factor $\beta > 0$ is chosen large enough to guarantee finiteness to the problem, and the set $\mathcal{A}(r)$ is the set of admissible controls  keeping the wealth nonnegative.

Let $\alpha_0\in[0,r]$ and define, in the random interval $[0,\tau_1)$, the processes $X,Y,J$ as
\beqs
dX_t\ =\  - c_t dt + \pi_t (b_L dt + \sigma_L dW_t), && X_0\ =\ r-\alpha_0, \\
dY_t \ =\  Y_t \Big( \frac{\rho b_L \sigma_I}{\sigma_L} dt + \rho \sigma_I dW_t\Big), && Y_0\ =\ \alpha_0,\\
J_t \ =\  \alpha_0\frac{I_t }{Y_t}.
\enqs
In other words, $X_t$ is the liquid wealth at time $t$ (the wealth held in the riskless or in the liquid asset),  $Y_tJ_t$ is the wealth held  in the illiquid asset $I$, and the total wealth is $R_t=X_t+Y_tJ_t$.

We may  apply a Dynamic Programming Principle between $0$ and $\tau_1$, and see that $V$ satisfies the following dynamic programming principle:
\beq \label{eqDPPli}
V(r) &=& \sup_{0 \leq \alpha_0 \leq r} \sup_{(c_t,\pi_t)\in\mathcal{A}'(r,\alpha_0)} \E \left[\int_0^{\tau_1} e^{- \beta s} U(c_s) ds + e^{- \beta \tau_1} V(R_{\tau_1}) \right],
\enq
where $\mathcal{A}'(r,\alpha_0)$ is the set of admissible controls $(c_t,\pi_t)$ keeping the process $X$  nonnegative in the interval $[0,\tau_1)$.
Let us focus on the inner optimization problem in \eqref{eqDPPli}, i.e. assume that $\alpha_0$ is fixed and we want to optimize only on $(c_t,\pi_t)\in\mathcal{A}'(r,\alpha_0)$, and let us show  how this problem may be rewritten so as to fall in the framework of Subsection \ref{subsecrandom}.

Let ${\mathbb{F}}^W=(\mathcal{F}^W)_{t\geq 0}$ denote the filtration generated by $W$. We note that $Y$ is ${\mathbb{F}}^W$-adapted, while $J$ is independent of ${\mathbb{F}}^W$. Moreover, since $I$ is not observed in the interval $[0,\tau_1)$, the information available to the investor is given by the filtration ${\mathbb{F}}^W$ in that interval.
Hence,
defining  the function $(t,x,y)\mapsto G[V](t,x,y):=\E[V(x+yJ_t)]$ and
taking the conditional expectation with respect to ${\cal F}^W_{\tau_1}$ in the inner optimization problem of \reff{eqDPPli}, this last one may be rewritten as
\beq \label{eqIPli}
\sup_{(c_t,\pi_t)\in\mathcal{A}'(r,\alpha_0)} \E \left[\int_0^{\tau_1} e^{- \beta s} U(c_s) ds + e^{- \beta \tau_1} G[V](\tau_1, X_{\tau_1}, Y_{\tau_1}) \right].
\enq
Now,
if we choose $U(c) = \frac{c^p}{p}$, $p$ $\in$ $(0,1)$, the value function $V$ will be $p$-homogeneous,  $V(r) = K_V \frac{r^p}{p}$, and we can reduce the state space of the above inner control problem to one space dimension. Indeed, let us consider the state variable $Z_t := \frac{X_t}{Y_t}$.
Letting
\beqs
 \tilde{c}_s \ =\  \frac{c_s}{Y_s},\  \ \ \ \ \
 \tilde{\theta}_s  \ =\ \frac{\pi_s}{Y_s} - Z_s \frac{\rho \sigma_I}{\sigma_L},
\enqs
one can check that $Z$ is a solution of the SDE
\beq \label{eqSDE-Zli}
dZ_t &=& - \tilde{c}_t dt + \tilde{\theta}_t \left((b_L - \rho \sigma_I \sigma_L) dt + \sigma_L dW_t\right), \ \ \ \ \ Z_0\ =\ z\ =\ \frac{r-\alpha_0}{\alpha_0}.
\enq
Furthermore, \reff{eqIPli} may be rewritten as
\beq\label{FFF}
\sup_{(\tilde c_t, \tilde \theta_t)\in \mathcal{A}''(z)} \E \left[\int_0^{\tau_1} e^{- \beta s} Y_s^p U(\tilde c_s) ds + e^{- \beta \tau_1} Y_{\tau_1}^p G[V](\tau_1, Z_{\tau_1}, 1) \right],\enq
where $\mathcal{A}''(z)$ is the set of admissible controls  $(\tilde c_t, \tilde \theta_t)$ keeping the process $Z$ nonnegative.
We can rewrite \eqref{FFF}
just in terms of $Z$. In order to do that, notice that $Y_t^p = \alpha_0^p H_t e^{k_{Y,p}t}$, where $k_{Y,p} = p \rho \frac{\sigma_I b_L}{\sigma_L} - \frac{p(1-p) \rho^2 \sigma_I^2}{2}$ and $H$ is a martingale defined by $H_0=1$, $dH_s = p\rho \sigma_I H_s dW_s$.
Then,
denoting by $\Q$ the probability with density process $H_t$, we have that $\widehat W_t := W_t - p \rho \sigma_I t$ is a $\Q$-Brownian motion. Moreover,
\reff{eqSDE-Zli} is equivalent to
\beq\label{stateZ}
dZ_t &=& - \tilde{c}_t dt + \tilde{\theta}_t \left((b_L - \rho \sigma_I \sigma_L(1-p)) dt + \sigma_L d\widehat W_t\right),
\enq
and the control problem can be rewritten as
\beq\label{funZ}
\alpha_0^p \cdot \sup_{(\tilde c, \tilde \theta)\in\mathcal{A}''(z)} \E^\Q \left[\int_0^{\tau_1} e^{- (\beta-k_{Y,p}) s} U(\tilde c_s) ds + e^{- (\beta-k_{Y,p}) \tau_1} G[V](\tau_1, Z_{\tau_1}, 1) \right].
\enq
Due to Subsection  \ref{subsecrandom}, the optimization problem \eqref{stateZ}-\eqref{funZ} is now in the framework of this paper (as long as we assume that $\tau_1$ has a density).


\end{document}